\documentclass[12pt, letterpaper]{article}

\usepackage{amsmath}
\usepackage{amsfonts}
\usepackage{amssymb}
\usepackage{graphicx}

\usepackage{color}

\usepackage{amsmath, amssymb, graphics, epsfig, graphicx}
\usepackage{epsf}
\usepackage{epstopdf}
\usepackage {amssymb}
\newcommand{\nc}{\newcommand}
\nc{\ba}{\begin{eqnarray}}
\nc{\ea}{\end{eqnarray}}
\newcommand\be{\begin{equation}}
\newcommand\ee{\end{equation}}

\begin{document}

\begin{flushright} {\footnotesize YITP-20-20}  \end{flushright}
	\vspace{0.5cm}

\begin{center}

\def\thefootnote{\fnsymbol{footnote}}

{\Large\bf Mimetic Black Holes}
\\[0.7cm]

{Mohammad Ali Gorji$^{1,2}$ \footnote{gorji@yukawa.kyoto-u.ac.jp}, 
Alireza Allahyari$^2$ \footnote{alireza.al@ipm.ir}, 
Mohsen Khodadi$^2$ \footnote{m.khodadi@ipm.ir},
Hassan Firouzjahi$^{2,3}$ \footnote{firouz@ipm.ir}
}
\\[0.5cm]
 
 {\small \textit{$^1$Center for Gravitational Physics, Yukawa Institute for Theoretical Physics, Kyoto University, \\ Kyoto 606-8502, Japan
}}\\
 
 {\small \textit{$^2$School of Astronomy, Institute for Research in Fundamental Sciences (IPM), \\ P.~O.~Box 19395-5531, Tehran, Iran
}}\\

{\small \textit{$^3$ Department of Physics, Faculty of Basic Sciences, University of Mazandaran, \\ 
P. O. Box 47416-95447, Babolsar, Iran}}\\

\end{center}

\vspace{.8cm}

\hrule \vspace{0.3cm}


\begin{abstract}
In this paper, we look for the vacuum static spherically symmetric solution in the mimetic gravity scenario based on the conformal invariance principle. The trivial solution is a stealth Schwarzschild black hole with scalar hair where the mimetic field does not contribute to the background. However, a  solution with two naked singularities shows up when the mimetic scalar field contributes to the background. We show that one of these singularities is the same as the singularity at the center of standard Schwarzschild black hole while the other appears due to caustics formation. However, we construct the mimetic black hole solution by gluing the exterior static spherically symmetric solution to a time-dependent anisotropic spacetime describing the interior of the black hole. It is shown that these two solutions match continuously on the surface of the apparent horizon. Some physical properties of the corresponding mimetic black holes are discussed. 
\end{abstract}
\vspace{0.5cm} \hrule
\def\thefootnote{\arabic{footnote}}
\setcounter{footnote}{0}


\newpage

\section{Introduction}

Mimetic gravity is a scalar-tensor gravity in which the conformal mode of gravity is isolated by means of a scalar field \cite{Chamseddine:2013kea}. Alternatively, the setup can be viewed as a special case of a general conformal/disformal transformation in which the transformation between the old and new metrics is degenerate. With a non-invertible conformal/disformal transformation the number of degrees of freedom can increase such that the longitudinal mode of gravity will become dynamical \cite{Deruelle:2014zza, Domenech:2015tca, Firouzjahi:2018xob,Shen:2019nyp}. The conformal transformation relating the physical metric $g_{\mu\nu}$ to the auxiliary metric ${\tilde g}_{\mu\nu}$ and the scalar field $\phi$ is given by\footnote{We work with the metric signature $(-,+,+,+)$.} 
\begin{equation}\label{mimetic-trans}
g_{\mu\nu} = \pm \big( {\tilde g}^{\alpha\beta}\partial_{\alpha}\phi\partial_{\beta}\phi \big)\, {\tilde g}_{\mu\nu} \,.
\end{equation}
Note that the physical metric $g_{\mu\nu}$ is invariant under conformal transformation of the auxiliary metric ${\tilde g}_{\mu\nu}$. This principle uniquely fixes the functional form of the conformal factor in terms of the auxiliary metric ${\tilde g}_{\mu\nu}$ and the scalar field $\phi$ while it cannot fix the overall sign. We have shown this fact in the appendix \ref{app-a}. The above transformation implies that the physical metric satisfies the following constraint
\ba\label{mimetic-const.}
g^{\mu \nu } \partial_\mu \phi \partial_\nu \phi  = \pm 1 \,.
\ea
Therefore, the vector $\partial^{\mu}\phi$ is spacelike for the $+$ sign while it is timelike for the $-$ sign. The $-$ sign in (\ref{mimetic-trans}) and (\ref{mimetic-const.}) corresponds to the standard mimetic gravity scenario \cite{Chamseddine:2013kea} while the $+$ sign can be thought as a generalization of the mimetic gravity.

The most important point about the mimetic transformation (\ref{mimetic-trans}) is that it is a singular transformation so that we cannot express the auxiliary metric ${\tilde g}_{\mu\nu}$ in terms of the physical metric $g_{\mu\nu}$ \cite{Deruelle:2014zza}. The resultant new degree of freedom associated with the transformation (\ref{mimetic-trans}) represents  the longitudinal mode of gravity which is dynamical even in the absence of any matter field. If we start with the vacuum Einstein-Hilbert action in terms of the physical metric $g_{\mu\nu}$ and then performing the mimetic transformation (\ref{mimetic-trans}), we end up with a scalar-tensor theory in terms of the auxiliary metric ${\tilde g}_{\mu\nu}$ and dynamical scalar field $\phi$ \cite{Chamseddine:2013kea}.  Equivalently, one can work with the physical metric $g_{\mu\nu}$ and dynamical scalar field $\phi$ by taking the constraint Eq.  (\ref{mimetic-const.}) into account by means of a Lagrange multiplier as follows \footnote{We work in units $M_{\rm P}=1/\sqrt{8\pi{G}}=1$.} \cite{Golovnev:2013jxa}
\ba\label{action}
S_{\pm}= \int d^4 x  \sqrt{-g} \left[\, \frac{R}{2} + \lambda \Big( g^{\mu \nu } \partial_\mu \phi \partial_\nu \phi \mp 1 \Big) \right] \,,
\ea
where $R$ is the Ricci scalar constructed from  the physical metric $g_{\mu\nu}$ while the auxiliary field $\lambda$ enforces the mimetic constraint Eq. (\ref{mimetic-const.}). We have allowed for the vector field $\partial^{\mu}\phi$ to be either spacelike or timelike which is  described by the actions $S_{+}$ and $S_{-}$  respectively.

We note that although there are two actions in our scenario in Eq. (\ref{action}), but we still deal with one theory. As we mentioned above the physical metric $g_{\mu\nu}$ in Eq. (\ref{mimetic-trans}) is invariant under conformal transformation of the auxiliary metric ${\tilde g}_{\mu\nu}$ for both cases of $\pm$. The scenario is based on this principle of conformal invariance which cannot fix the $\pm$ signs \cite{Deruelle:2014zza}. To find a cosmological solution one chooses to work with the $-$ sign and a time-dependent ansatz for the scalar field which is consistent with the homogeneity of the background under consideration. Since the interior of a static black hole is also time-dependent, one similarly needs to work with the $-$ sign as we will see in the following sections where  we study the interior solution of the mimetic black hole. In the case of static solution, as we will show in the next section, we have to consider the $+$ sign.  Here, isolating the conformal degrees of freedom based on the principle of conformal invariance leads to two different classes of actions. Depending on the system under consideration we can use whichever that is consistent. To find a black hole solution we need to glue the exterior and interior solutions which are solutions of different actions but in the context of the same theory. In this regard, the situation is somewhat different than general relativity where one implements the standard junction condition to glue two different solutions to each other. Here, in addition to the standard junction conditions, we also check that the energy-momentum tensor of the mimetic sectors in both sides match 
smoothly and consistently to each other at the matching surface. In this manner, we can obtain a static black hole solution in this scenario.

The cosmological implications of the mimetic setup with the action $S_{-}$ and the timelike vector $\partial^\mu\phi$ are widely studied in recent years \cite{mimetic,Sadeghnezhad:2017hmr,Liu:2017puc,Dutta:2017fjw,Casalino:2018tcd,Brahma:2018dwx,deHaro:2018sqw,deCesare:2018cts,Ganz:2019vre,Myrzakulov:2015qaa}. Interestingly, an energy density component naturally arises in this scenario which behaves like dark matter \cite{Chamseddine:2013kea}.  Indeed, the action (\ref{action}) can be viewed as standard general relativity with dust as matter source in the  formalism of \cite{Schutz:1970my,Brown:1992kc} so that $\lambda$ and $\phi$ play the roles of energy density and velocity potential respectively. In principle, however, the mimetic theory is different than general relativity. In mimetic gravity, both metric and $\phi$ are fundamental fields while in general relativity, only metric is the fundamental field. Although the two setups are the same in practice, it is important to note that they are fundamentally different. For instance, due to the attractive nature of the resultant dark matter, both the mimetic dark matter scenario and general relativity with dust as matter source would finally end up with caustic singularities around them the field $\phi$ is ill-defined. This is a real pathology for the mimetic scenario as already mentioned in the literature \cite{Barvinsky:2013mea,mukohyama-caustics,caustics}. However, it is not a real problem for general relativity with dust as matter source since one may simply claim that the fluid description in the matter sector breaks down near caustics which is not an issue for the gravitational sector. To solve this problem in mimetic scenario, gauge field extension of the mimetic scenario is proposed in which the scalar field is replaced with a gauge field and the setup is free of caustic singularities \cite{Vikman:2017gxs,Gorji:2018okn,Jirousek:2018ago,Gorji:2019ttx}. Moreover, the sound speed for the scalar perturbations in mimetic dark matter scenario vanishes and in order to generate a propagating mode, it is suggested to add some higher derivative terms of the mimetic scalar field to the action \cite{Chamseddine:2014vna,Mirzagholi:2014ifa}. The resultant setup then suffers from the ghost and/or gradient instabilities \cite{Ramazanov:2016xhp,Ijjas:2016pad,Firouzjahi:2017txv} and one needs to add a non-minimal coupling between the higher derivatives of the mimetic scalar field and the curvature terms  to make the setup stable \cite{Hirano:2017zox,Zheng:2017qfs,Gorji:2017cai,Langlois:2018jdg}.

In this paper, we construct  the black hole solution in the original mimetic scenario defined by the action Eq.~(\ref{action}). Some attempts have been made  in this direction in the Refs. \cite{Myrzakulov:2015kda,Oikonomou:2016fxb,Chen:2017ify,Li:2018uwg,Nashed:2018qag,Sheykhi:2019gvk}. But the difficult part is that in order to have a static black hole solution we need to have a horizon. As we will show, obtaining a horizon for a static black hole solution in the mimetic gravity scenario is not an easy task. More precisely, unlike in cosmological setups where one uses only the timelike vector $\partial^\mu\phi$, here  we need both timelike and spacelike vectors $\partial^\mu\phi$.

\section{Static Solutions}\label{ext}

In this section, our aim is to find a static spherically symmetric solution in the mimetic gravity scenario. 
We show that the static spherically symmetric solution  is not a black hole solution but a spacetime with naked singularities. We present a black hole solution with a proper horizon in next Section. 

The general scalar field and the the Einstein field equations are 
\ba
\label{KG-Eqs}
\partial_\mu \big( \sqrt{-g} \lambda g^{\mu\nu}\partial_\nu \phi \big) = 0 \,,
\ea
and
\ba
\label{Einstein-Eqs}
R_{\mu\nu } - \frac{1}{2} R g_{\mu\nu} = - 2 \lambda \partial_\mu \phi \partial_\nu\phi  \,,
\ea
where the mimetic constraint Eq. (\ref{mimetic-const.}) has been imposed to simplify the last equation.

The metric in local coordinates $(t,r,\theta,\varphi)$ has the form
\begin{eqnarray}\label{metric-schw}
ds^2 = -A(r) dt^2 + \frac{dr^2}{B(r)} + r^2 d\Omega^2 \,,
\end{eqnarray}
where $A(r)>0$ and $B(r)>0$ are two unknown functions which should be determined after solving the corresponding dynamical equations and $d\Omega^2 = d\theta^2+\sin^2\theta \, d\varphi^2$ is the metric of a sphere with unit radius.

The static metric ansatz (\ref{metric-schw}) does not necessarily imply static configuration (\ref{phi-lambda-ansatz-r}) and we consider the following general configuration
\begin{eqnarray}\label{phi-lambda-ansatz-tr}
\phi=\phi(t,r) \,, \hspace{1cm} \lambda=\lambda(t,r) \,.
\end{eqnarray}

From the off-diagonal $tr$ component of the Einstein equations associated to the action (\ref{action}), we find
\begin{equation}\label{EE-tr}
\mp \lambda \partial_t\phi \partial_r \phi = 0 \,,
\end{equation}
where $-$ and $+$ signs correspond to $S_{+}$ and $S_{-}$ respectively. Substituting from (\ref{metric-schw}) and (\ref{phi-lambda-ansatz-tr}), the mimetic constraint (\ref{mimetic-const.}) yields 

\ba\label{mimetic-const-general}
- \frac{1}{A} (\partial_t \phi)^2 + B (\partial_r \phi)^2  = \pm 1 \,.
\ea

In order to satisfy Eq (\ref{EE-tr}), we can choose either $\lambda (t,r) = 0$ or $\lambda (t,r) \neq 0$. The mimetic scalar field does not contribute to the background in the first while it contributes to the background for the latter choice. In the following subsections, we study both possibilities in details.

\subsection{Stealth Schwarzschild Black Holes}
If we set $\lambda=0$, the energy-momentum of mimetic scalar field vanishes and all the mimetic effects disappear in the Einstein equations (\ref{Einstein-Eqs}) and we end up with well known solution
\begin{equation}\label{stealth-AB}
A(r) = B(r) = 1 - \frac{2m}{r},
\end{equation}
where $m$ is the mass of the black hole. Substituting the above solution back into the mimetic constraint (\ref{mimetic-const-general}) gives
\ba\label{mimetic-const-stealth}
- \frac{r}{r-2m} (\partial_t \phi)^2 + \frac{r-2m}{r} (\partial_r \phi)^2  = \pm 1 \,,
\ea
which can be solved to find solutions for the mimetic scalar field $\phi$. Therefore, although background metric is the same as standard Schwarzschild black hole, the mimetic field propagates on top of this background as scalar hair. These are stealth Schwarzschild solutions which usually show up in scalar-tensor theories \cite{Herdeiro:2014goa,Sotiriou:2015pka}. In recent years, stealth black hole solutions are widely studied in the context of higher derivative scalar-tensor theories \cite{Babichev:2013cya,Sotiriou:2014pfa,Babichev:2016rlq,Minamitsuji:2018vuw,BenAchour:2018dap,BenAchour:2019fdf,Motohashi:2019ymr}. To solve (\ref{mimetic-const-stealth}), we note that $\partial_{\mu}\phi$ satisfies geodesic equation which can be seen from the mimetic constraint (\ref{mimetic-const.}). Therefore, choosing $\phi$ as proper time in the case of $S_-$, we can find many solutions through the evolution of congruence of $\partial_{\mu}\phi$ with different initial conditions. The same argument can be applied to the case of $S_+$. For instance, considering ansatz with linear time dependence results in $\phi = t \pm \int \big(\frac{\sqrt{1\mp{A}}}{A}\big) dr$ where $\mp$ signs in the integral correspond to $S_-$ and $S_+$ respectively. This solution diverges at the horizon $r = 2m$ while the horizon limit in ingoing Eddington-Finkelstein coordinates is well-defined. In this regard, we can also find stealth charged Reissner-Nordstrom black hole with scalar hair as well. Existence of these general relativity solutions is interesting since it shows that the mimetic scenario includes the standard black hole solutions at least as stealth solution.

Thus, if the mimetic scalar field does not contribute to the background, mimetic scenario supports stealth Schwarzschild solutions. This is a new solution because of the scalar hairs while the role of mimetic scalar field completely ignored at level of background. Therefore, it is even more interesting if we consider the case $\lambda\neq0$ when the mimetic scalar field contributes to the background. The solution then will be no longer stealth and different than Schwarzschild black hole. This is the subject of the next subsection and, in a sense, it is a counterpart of mimetic dark matter where the mimetic field plays the role of dark matter in the cosmological background with $\lambda\neq0$.

\subsection{Static Solution with Naked Singularities}

As soon as we take $\lambda\neq0$, the off-diagonal $tr$ component of the Einstein equations (\ref{Einstein-Eqs}) implies that we have to choose either $\partial_t\phi=0$ or $\partial_r \phi=0$ for both $S_{\pm}$. In the case of static background (\ref{metric-schw}), we conclude that $\partial_t\phi =0$. Then, from the mimetic constraint (\ref{mimetic-const-general}) we find that only $S_{+}$ is consistent with the static ansatz since $S_{-}$ leads to an imaginary scalar field. Note that in the case of stealth solution with $\lambda=0$, both scenarios $S_{\pm}$ were acceptable while we have to only work with $S_{+}$ for $\lambda\neq0$. This is in agreement with the no-go result that is obtained in the context of Horava-Lifshitz gravity in \cite{mukohyama-nogo}. Indeed, mimetic model with timelike vector field $\partial^\mu\phi$ defined by $S_{-}$ also emerges from the infrared (IR) limit of Horava-Lifshitz gravity and  Einstein-aether scenario \cite{Blas:2009yd,Barausse:2011pu}. Thus, we end up with the following static configuration in this case
\begin{eqnarray}\label{phi-lambda-ansatz-r}
\phi=\phi(r) \,, \hspace{1cm} \lambda=\lambda(r) \,.
\end{eqnarray}

Now, substituting from Eqs. (\ref{metric-schw}) and (\ref{phi-lambda-ansatz-r}) in mimetic constraint (\ref{mimetic-const.}), we find
\begin{eqnarray}\label{mimetic-constarint}
B (\partial_{r}\phi)^2 = 1 \,,
\end{eqnarray}
which yields
\begin{eqnarray}\label{phiprime}
\phi' = \pm \frac{1}{\sqrt{B}} \,.
\end{eqnarray}
Without loss of generality we take the positive branch of solution in our forthcoming analysis.

The $tt$ component of the Einstein equations (\ref{Einstein-Eqs}) yields
\begin{equation}\label{Einstein-tt}
\partial_r (r B) = 1 \,,
\end{equation}
where we have substituted from Eq. (\ref{phiprime}). Solving the above equation yields
\begin{eqnarray}\label{B-0}
B = 1 + \frac{c_0}{r} \,,
\end{eqnarray}
where $c_0$ is an integration constant. At this point we cannot determine the sign of the constant $c_0$. Indeed, we will show that it can be either positive or negative. In this subsection, we focus on the positive case and we will come back to the case of $c_0<0$ to construct a black hole solution in the next section.

Before continuing with the other components of the  Einstein equations, it is helpful to look at the mimetic field equation (\ref{KG-Eqs}) which in coordinates defined in Eq.~(\ref{metric-schw}) yields
\begin{equation}
\partial_r(\sqrt{A}r^2\, \lambda) = 0 \,,
\end{equation}
where we have used the fact that $\sqrt{-g}=\sqrt{A/B}r^2\sin\theta$ and also substituted from Eq.~(\ref{phiprime}). The above equation can be solved to give
\begin{equation}\label{A-lambda}
A = \Big( \frac{c_1}{r^2\lambda} \Big)^2 \,,
\end{equation}
where $c_1$ is another integration constant. Note that $A$ is always positive and, therefore, we would have $B>0$. This confirms that the static solution can only be achieved in $S_{+}$ model.

Substituting Eqs. (\ref{phiprime}), (\ref{B-0}) and (\ref{A-lambda})  in the $rr$ component of the Einstein equation, we find the following differential equation for the Lagrange multiplier $\lambda$, 
\begin{eqnarray}\label{Einstein-rr}
(r+c_0) \frac{\lambda'}{\lambda}+r^2\lambda+ 2 + \frac{3c_0}{2r} = 0 \,.
\end{eqnarray}
The above equation can be easily integrated yielding the following solution
\begin{eqnarray}\label{lambda}
\lambda = \frac{1}{2 r^2} \Big[ 1 - \sqrt{1+c_0/r} \ln\Big(\sqrt{r/c_2}\big(1+\sqrt{1+c_0/r} \,\big)\Big)\, \Big]^{-1} \,,
\end{eqnarray}
where $c_2$ is a constant of integration. From the above solution, we see that $\lambda$ diverges at two points $r=0$ and $r=r_f$ which solves the equation $1 = \sqrt{1+c_0/r_f} \ln\Big(\sqrt{r_f/c_2}\big(1+\sqrt{1+c_0/r_f} \,\big)\Big)$. Solving this equation for $c_2$ and then substituting the result back in Eq. (\ref{lambda}), we find
\begin{eqnarray}\label{lambda-rf}
\lambda = \frac{1}{2 r^2} \bigg[ 1 - \sqrt{1+c_0/r} \bigg(\frac{1}{\sqrt{1+c_0/r_f}} + \ln\sqrt{\frac{r}{r_f}}\frac{1+\sqrt{1+c_0/r}}{1+\sqrt{1+c_0/r_f}}\,\bigg)\, \bigg]^{-1} \,.
\end{eqnarray}

\begin{figure}
	\begin{center}
		\includegraphics[scale=0.45]{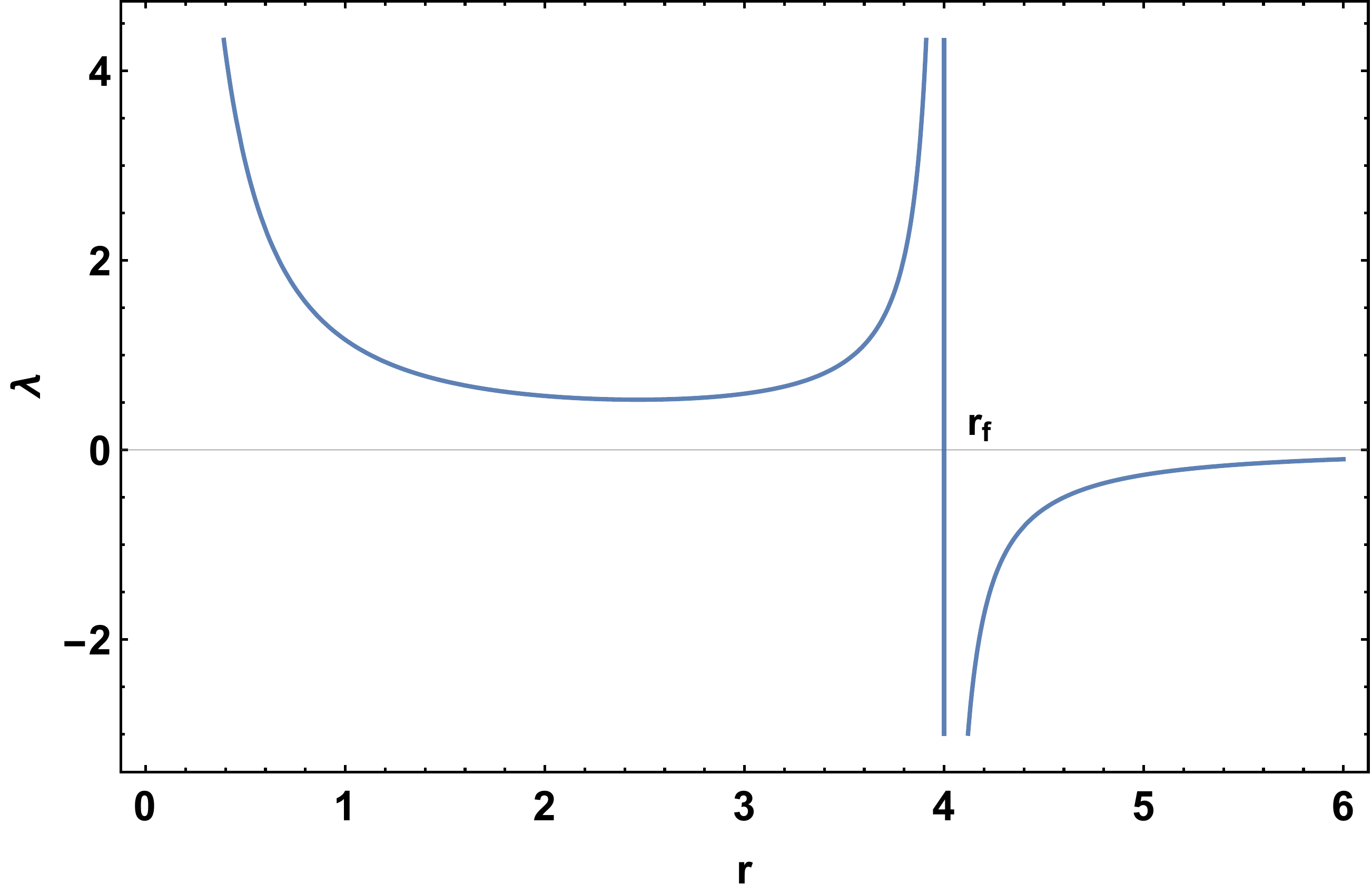}
		\caption{\label{fig1} The plot of $\lambda$ given in Eq. (\ref{lambda-rf}) versus the radial coordinate $r$  for $c_0=2$ and $r_f=4$. Since $\lambda$ also characterizes the curvature invariants like Ricci scalar, we see that there is a singularity at $r=r_f$ after which the curvature changes sign. We show that $r=r_f$ is a caustic singularity while $r=0$ is the usual curvature singularity.}	
	\end{center}	
\end{figure}

The behaviour of $\lambda$ as a function of $r$ is plotted in Fig. \ref{fig1} which shows that there is a branch cut at $r=r_f$ so that $\lambda$ changes sign with $\lim\lambda_{r\to {r_{f}^{-}}} = +\infty$ and $\lim\lambda_{r\to {r_{f}^{+}}} = -\infty$.

To understand whether these divergences are genuine  or the  coordinate artifacts, we look at the curvature invariants. In the case of Schwarzschild black hole, the Ricci scalar and Ricci tensor both vanish so in order to investigate the singularity properties, one usually looks at the Kretschmann scalar $R_{\mu\nu\alpha\beta}R^{\mu\nu\alpha\beta}$. In our setup, however, the Ricci scalar and Ricci tensor are both non-zero so it is reasonable to first look at the scalar quantities constructed from them. Taking the trace of the Einstein equation (\ref{Einstein-Eqs}) and using the mimetic constraint Eq. (\ref{mimetic-const.}) for the spacelike case, we find
\begin{equation}\label{ricci-scalar}
R = 2 \lambda \,.
\end{equation}
Substituting the above relation back into the Einstein equation (\ref{Einstein-Eqs}), we find
\begin{equation}\label{ricci-tensor}
R_{\mu\nu} = \lambda g_{\mu\nu} - 2 \lambda \partial_\mu\phi \partial_\nu\phi \,.
\end{equation}
In addition to the Ricci scalar which is given by (\ref{ricci-scalar}), we can construct another scalar quantity from the Ricci tensor with the same dimension as follows
\begin{equation}\label{ricci-tensor-phi}
\partial^\mu\phi\, \partial^\nu\phi R_{\mu\nu}  = \lambda \,.
\end{equation}
The next nonzero scalar curvature which can be constructed is $R_{\mu\nu} R^{\mu\nu} $, which after using the  mimetic constraint Eq. (\ref{mimetic-const.}) in Eq. (\ref{ricci-tensor}) yields 
\begin{equation}\label{ricci-tensor-2}
R_{\mu\nu} R^{\mu\nu}  = 4 \lambda^2 \,.
\end{equation}
We can also write the Kretschmann scalar in terms of $\lambda$ as
\begin{equation}\label{kretschmann}
R_{\mu\nu\alpha\beta}R^{\mu\nu\alpha\beta} =
12 \lambda ^2-\frac{16 \lambda  c_0}{r^3} +\frac{12 c_0^2}{r^6} \,,
\end{equation}
while the Weyl squared scalar is given by
\begin{equation}\label{weyl2}
C_{\mu\nu\alpha\beta} C^{\mu\nu\alpha\beta} = \frac{16}{3} \left(\lambda - \frac{3c_0}{2r^3} \right)^2 \,.
\end{equation}

From the above relations we see that $\lambda$ characterizes all curvature invariants and therefore the divergences at $r=0$ and $r=r_f$ ( see Fig. \ref{fig1}) are real singularities in this scenario. Since there is 
no horizon in this solution, $r=0$ and $r\to{r_f}$ are {\it naked singularities}. This is the the natural static spherically symmetric solution of the mimetic scenario.

If we only look at the curvature invariants, we cannot distinct between the two singularities $r=0$ and $r=r_f$. It is then useful to look at $g_{tt} = - (4 r^4 \lambda^2)^{-1}$. From Eq. (\ref{lambda-rf}), it is clear that $g_{tt}$ vanishes at $r_f$ but not at $r=0$. In Section \ref{singularities} we will show that indeed the types of singularities at $r=0$ and $r=r_f$ are totally different so that the first is the usual curvature singularity which also arises in the standard general relativity (much similar to what arises in the center of standard black hole solutions like Schwarzschild solution) while the latter is a caustic-type singularity which is the characteristic  of the mimetic gravity scenario.

From Eq. (\ref{ricci-scalar}) and Fig \ref{fig1}, we see that the two regions $r\in[0,r_f]$ and $r\in[r_f,\infty]$ describe regions with positive and negative curvatures respectively. We therefore take the positive region as our solution and choose $r_f$ to be sufficiently large as $r_f\to\infty$ to exclude the region with negative curvature.

It is straightforward to show that all other Einstein's equations are satisfied and our solution is complete now. Substituting from Eqs.~(\ref{B-0}) and (\ref{A-lambda}), the metric  Eq.~(\ref{metric-schw}) takes the following form
\begin{eqnarray}\label{metric-naked-sing}
ds^2 = - \frac{dt^2}{4 r^4 \lambda^2} + \frac{dr^2}{1+\frac{c_0}{r}} + r^2 d\Omega^2 \,,
\end{eqnarray}
where $\lambda$ is given by Eq.~(\ref{lambda-rf}). Now, substituting Eq. (\ref{lambda-rf}) in Eq. (\ref{A-lambda}), we find an explicit solution for $g_{tt}$. We see that the effects of the constant $c_1$ can be absorbed through rescaling the time via $t\to{t/c_1}$ in metric (\ref{metric-schw}) so on can set $c_1=1$ without loss of generality.

From Eqs. (\ref{mimetic-constarint}) and (\ref{A-lambda}), we also see that $g_{rr}$ and $g_{tt}$ cannot change sign and, therefore, our solution (\ref{metric-naked-sing}) does not have any horizon. Consequently,  our  solution (\ref{metric-naked-sing}) does not describe a black hole. To get some intuition, it is useful to compare our solution with its standard counterpart, {\it i.e.}, the Schwarzschild solution. In Schwarzschild solution, $g_{tt}=-g^{rr} = - ( 1 + \frac{c_0}{r})$ so that $g_{tt}$ is related to the Newtonian potential $\Phi$ in the weak field limit as $g_{tt} = -(1+2\Phi)$. We then conclude that $c_0$ should be negative which immediately implies the existence of a horizon since $g_{tt}=-g^{rr}$. In our solution (\ref{metric-naked-sing}) with $g_{tt}\neq-g^{rr}$, however, this is not the case. The reason is simple: contrary to GR the longitudinal mode of gravity encoded in the mimetic scalar field is dynamical here and has a non-zero profile in the whole spacetime. In other words, the energy of the mimetic scalar field is not localized and we cannot treat solution (\ref{metric-naked-sing}) as a gravitational field of an isolated source. Therefore, $c_0$ can be either positive or negative as we claimed above.

In the case of $c_0>0$ from Eq. (\ref{B-0}) we find that the condition $B>0$ required by the mimetic constraint (\ref{mimetic-constarint}) is satisfied for $r\in[0,r_f)$. Therefore, the metric (\ref{metric-naked-sing}) is applicable for the whole range of $r\in[0,r_f)$ in this case.

Substituting Eq. (\ref{B-0}) in Eq. (\ref{phiprime}) and then integrating, we find the following solution for the mimetic scalar field
\begin{equation}\label{phi}
\phi =  r \sqrt{1+c_0/r} - {c_0} \ln\Big(\sqrt{r/c_3}\big(1+\sqrt{1+c_0/r} \,\big)\Big) \,  \,,
\end{equation}
where $c_3$ is an integration constant which does not play any important role since our setup is shift-symmetric.

Finally, it is useful to look at the special case of $c_0=0$ that corresponds to the Minkowski spacetime in the case of GR with the Schwarzschild solution. From (\ref{ricci-scalar}) and (\ref{kretschmann}), we see that the curvature invariants do not vanish even if we set $c_0=0$. Then, it is reasonable to consider the effects of $c_0$  coming from a local matter energy density while $\lambda$ labels the  effects coming from the energy density induced by mimetic field. In other words, we have a dynamical degrees of freedom even in the absence of any local matter energy density. Therefore, we do not have a Minkowski spacetime even far from a local matter profile. To see this fact explicitly, we set $c_0=0$ in our previous results in which the metric (\ref{metric-naked-sing}) takes the following simple form
\begin{equation}\label{metric-c=0}
ds^2 = - \Big( \ln{\frac{r}{r_f}} \Big)^2 dt^2 + dr^2 + r^2 d\Omega^2 \,.
\end{equation}
The above result shows that the metric is not asymptotically flat in the allowed region of $r\in[0,r_f)$. The above metric is not singular around $r=0$ which shows that this singularity comes from the existence of a matter profile and disappears when we set $c_0=0$. The singularity at $r=r_f$ is, however, present even in the absence of any local matter field. This shows that the singularity at $r=r_f$ is the characteristic of the mimetic theory itself.
We will show that the mimetic scenario approaches a caustic singularity at $r = r_f$ and we cannot trust the model near this point.  

\section{Black Hole Solution}

In the previous Section we have shown that a static solution with naked singularities appears 
when the mimetic field contributes to the background. However, we only studied the case 
$c_0>0$ in Eq. (\ref{B-0}) for $g_{rr}=B^{-1}$. The situation is different if we consider $c_0<0$ in which the requirement $B>0$ from the mimetic constraint Eq. (\ref{mimetic-constarint}) implies that $r>|c_0|$. In the following subsections we study this case in details.

\subsection{Exterior Static Spherically Symmetric Spacetime}
\label{ext-BH}

To distinguish this solution with the metric (\ref{metric-naked-sing}), we work with the coordinates $(t_+,r_+,\theta,\varphi)$ in which the metric takes the form
\begin{eqnarray}\label{metric-schw-p}
ds_+^2 = -A_+(r_+) dt_+^2 + \frac{dr_+^2}{B_+(r_+)} + r_+^2 d\Omega^2 \,.
\end{eqnarray}
We did not change the angular coordinates $(\theta,\varphi)$ since the changes are only for the temporal and radial coordinates. Moreover, we denote all functions with lower index $+$  in this subsection.

For $c_0<0$, the solution (\ref{B-0}) can be written as
\begin{eqnarray}\label{B-p}
B_+ = 1 - \frac{2m_+}{r_+} \,,
\end{eqnarray}
where we have defined $c_0=-2m_{+}$ with $m_+>0$ to be a positive constant. The spacetime then has a horizon defined by $g^{r_+ r_+}(r_{+H})=0$ with
\begin{equation}\label{horizon-p}
r_{+H} = 2m_+ \,.
\end{equation}
We note that the above solution is only applicable for $r_+>2m_+$ since $B_+>0$ for the spacelike 
vector  $\partial^\mu\phi_+$.

All the solutions in the previous section are now applicable but restricted to the range $r_+>2m_+$. The metric takes the same form as Eq. (\ref{metric-naked-sing}) and
\begin{eqnarray}\label{metric-p}
ds_+^2 = - \frac{dt_+^2}{4 r_+^4 \lambda_+^2} + \frac{dr_+^2}{1-\frac{2m_+}{r_+}} + r_+^2 d\Omega^2 \,,
\end{eqnarray}
where $\lambda_+$ is given by (\ref{lambda-rf}) but replacing $c_0$ with $-2m_+$,
\begin{eqnarray}\label{lambda-rf-p}
\lambda_+ = \frac{1}{2 r_+^2} \Bigg[ 1 - \sqrt{1-\frac{2m_+}{r_+}} \Bigg(\frac{1}{\sqrt{1-\frac{2m_+}{r_f}}} + \ln\sqrt{\frac{r_+}{r_f}}\,\frac{1+\sqrt{1-\frac{2m_+}{r_+}}}{1+\sqrt{1-\frac{2m_+}{r_f}}}\,\Bigg) \Bigg]^{-1} \,.
\end{eqnarray}
The radial coordinate is restricted to the interval $r_+\in[2m_+,r_f)$ where we also have fixed the constant $c_1$ in (\ref{A-lambda}) such that $g_{t_+,t_+}(r_{+H}) = - 4 c_1^2  = -1$. This normalization does not affect the physical properties of the model since it only corresponds to rescaling the time via $t_+\to{t_+/c_1}$.

The solution (\ref{metric-p}) that we have found in this subsection has horiozn (\ref{horizon-p}) while the radial coordinate is restricted to the range $r_+\in[2m_+,r_f)$. Therefore this horizon cannot be an event horiozn. In order to understand the type of this horizon, we look at the expansion of the outgoing null geodesics orthogonal to a closed, two-dimensional, spatial hypersurface inside the timelike hypersurface $r_{+H} = 2m_+$ which becomes proportional to $\sqrt{B}$ \cite{Baumgarte:2010ndz} and vanishes at $r_{+H} = 2m_+$. Therefore, the surface $r_{+H} = 2m_+$ may be interpreted as  an apparent horiozn.
 
The metric (\ref{metric-p}) can be thought as the exterior solution of a black hole with apparent horizon $r_{+H}$ defined in Eq. (\ref{horizon-p}). To have a complete black hole solution, however, we need to find the interior solution as well. In the case of the Schwarzschild black hole in GR, the interior solution can be obtained by exchanging $t$ and $r$ coordinates in the exterior solution. More precisely, the interior solution can be obtained by solving Einstein's equation for a homogeneous but anisotropic background geometry. Here, we have to do the same. From the exterior metric (\ref{metric-p}), we see that by switching $t_+\rightarrow{r_-}$ and $r_+\rightarrow{t_-}$, we obtain a homogeneous but anisotropic metric for the interior solution.  Considering a time-dependent anisotropic metric together with a time-dependent scalar $\phi(t)$  and auxiliary field $\lambda(t)$, it is easy to show that the spacelike action $S_{+}$ is not consistent with the mimetic constraint (\ref{mimetic-const.}) and we have to use the timelike action $S_{-}$. In the next subsection, we  show that the action $S_{-}$ admits a time-dependent solution that matches the exterior solutions (\ref{metric-p}) and (\ref{lambda-rf-p}) on the null hypersurface of the apparent horizon.

\subsection{Interior Solution}
\label{int}

In this section, we focus on the timelike mimetic action $S_{-}$  to find the interior solution in extension to   the exterior solution  obtained in the previous subsection. Working in local coordinates $(t_-,r_-,\theta,\varphi)$, we consider the following homogeneous but anisotropic background
\begin{eqnarray}\label{metric-schw-n}
ds_-^2 = - \frac{dt_-^2}{B_-(t_-)} + A_-(t_-) dr_-^2 + t_-^2 d\Omega^2 \,,
\end{eqnarray}
where $B_-$ and $A_-$ are two unknown functions of $t_-$. We also consider the following time-dependent ansatz for the mimetic sector
\begin{eqnarray}\label{phi-lambda-t}
\phi_- = \phi_-(t_-) \,, \hspace{1cm} \lambda_- = \lambda_-(t_-) \,,
\end{eqnarray}
which are consistent with the geometry (\ref{metric-schw-n}). The mimetic constraint (\ref{mimetic-const.}) for the timelike vector $\partial^\mu\phi$ then results in
\begin{eqnarray}\label{mimetic-constarint-n}
(\partial_{t_-}{\phi}_-)^2 = \frac{1}{B_-} \,,
\end{eqnarray}
where a dot denotes the derivative with respect to $t_-$. Note that  had we worked with the spacelike model $S_{+}$, we would have found a minus sign in the right hand side of the above relation which was inconsistent with the assumption $B_->0$ in (\ref{metric-schw-n}). This is the main reason why we should employ the timelike action $S_-$ instead of the spacelike action $S_+$ when looking for the interior solution of the black hole. 

Solving Eq. (\ref{mimetic-constarint-n}), we find
\begin{eqnarray}\label{phiprime-n}
\partial_{t_-}{\phi}_- = \frac{1}{\sqrt{B_-}} \,,
\end{eqnarray}
where, without loss of generality, we took the positive branch in the above solution. The $r_-r_-$ component of the Einstein equations (\ref{Einstein-Eqs}) gives
\begin{equation}\label{Einstein-rr-n}
\partial_{t_-} (t_- B_-) = -1 \,,
\end{equation}
where the substitution from Eq. (\ref{phiprime-n}) has been made.  Solving the above equation gives
\begin{eqnarray}\label{B-n}
B_- = \frac{2 m_-}{t_-} - 1 \,,
\end{eqnarray}
where $m_-$ is an integration constant. Here, $m_-$ can be either positive or negative. We, however, are interested in an interior black hole solution so we take $m_->0$ which admits  a horizon 
$B_-(t_{-H})=0$ at
\begin{equation}\label{horizon-n}
t_{-H} = 2m_- \,.
\end{equation}

The mimetic scalar field equation (\ref{KG-Eqs}) gives
\begin{equation}
\partial_{t_-} (\sqrt{A_-}\, t_-^2\, \lambda_-) = 0 \,,
\end{equation}
where again a substitution from Eq. (\ref{phiprime-n}) has been made. The above equation yields
\begin{equation}\label{A-lambda-n}
A_- = \Big( \frac{{\tilde c}_1}{t_-^2 \lambda_-} \Big)^2 \,,
\end{equation}
where ${\tilde c}_{1}$ is an integration constant. 

Substituting from Eqs. (\ref{phiprime-n}), (\ref{B-n}) and (\ref{A-lambda-n}) in the $t_-t_-$ component of the Einstein equations (\ref{Einstein-Eqs}) gives
\begin{eqnarray}\label{Einstein-tt-n}
(t_--2m_-) \frac{\partial_{t_-}{\lambda}_-}{\lambda_-}+t_-^2\lambda_- + 2 - \frac{3 m_-}{ t_- } = 0 \,.
\end{eqnarray}
The above first order differential equation can be integrated to give the following solution
\begin{eqnarray}\label{lambda-n}
\lambda_- = - \frac{1}{2 t_-^2} \Bigg[ 1 - \sqrt{\frac{2m_-}{t_-} - 1} \Bigg( \arctan \sqrt{\frac{2m_-}{t_-} - 1}+ {\tilde c}_2 \Bigg)\, \Bigg]^{-1} \,,
\end{eqnarray}
where ${\tilde c}_2$ is another constant of integration. The above solution shows that $\lambda_{-}$ diverges at two times $t_{-}=0$ and $t_{-}=t_{i}$ which is defined via $1 = \sqrt{\frac{2m_-}{t_i-} - 1} \Bigg( \arctan \sqrt{\frac{2m_-}{t_i} - 1}+ {\tilde c}_2 \Bigg)$. Solving this equation for ${\tilde c}_2$ and substituting the result back into the above solution, we find
\begin{eqnarray}\label{lambda-tf}
\lambda_- =  -\frac{1}{2 t_-^2} \Bigg[ 1-\sqrt{\frac{2m_-}{t_-} - 1} \Bigg(\frac{1}{\sqrt{\frac{2m_-}{t_i} - 1}}+\arctan \sqrt{\frac{2m_-}{t_-} - 1}-\arctan \sqrt{\frac{2m_-}{t_i} - 1}\,\Bigg)\, \Bigg]^{-1} .
\end{eqnarray}
Much similar to the discussion after Eq. (\ref{lambda-rf}), we can easily show that all curvature invariants diverge at the times $t_{-}=0$ and $t_{-}=t_{i}$ and the region $t_{-}\in[0,t_i)$ corresponds to the negative curvature. We will discuss these singularities in details in Section \ref{singularities}.

Substituting Eq. (\ref{lambda-tf}) into Eq. (\ref{A-lambda-n}), we see that the effects of constant ${\tilde c}_1$ can be removed in  the metric (\ref{metric-schw-n}) through re-scaling the radius via $r_-\to r_-/{\tilde c}_1$. However, note that $A_-(t_{-H})= 4 {\tilde c}_1^2$ so we set ${\tilde c}_1=1/2$ and $A_-(t_{-H})=1$. This is just a normalization and does not affect the physical properties of our solution.

Our solutions is complete for all dynamical fields and the metric (\ref{metric-schw-n}) takes the following  form
\begin{eqnarray}\label{metric-n}
ds_-^2 = - \frac{d t_-^2}{\frac{2 m_-}{t_-}-1} + \frac{dr_-^2}{4 t_-^4 \lambda_-^2} + t_-^2 d\Omega^2 \,,
\end{eqnarray}
where the explicit form of $\lambda_-$ is given by Eq. (\ref{lambda-tf}). Note that the above metric is only applicable for $t_-\in[t_i,2 m_-]$.

The homogeneous and {\it isotropic} solution of mimetic scenario in FLRW cosmological background is known as mimetic dark matter  \cite{Chamseddine:2013kea}. In order to compare the above homogeneous but {\it anisotropic} solution with its FLRW isotropic counterpart, in Appendix \ref{appendix} we have presented our  anisotropic solution in another chart  which is more relevant for the cosmological purposes. The results resemble the isotropic cosmological solution in certain limits.

\subsection{Junction Conditions: Gluing Exterior and Interior Solutions}

We have found two different mimetic solutions in subsections \ref{ext-BH} and \ref{int}. Our aim in this subsection is to check whether we can match the static spherically symmetric solution (\ref{metric-p}) with the homogeneous anisotropic time-dependent solution (\ref{metric-n}) as the exterior and interior of the mimetic black hole, respectively. In order to do this, we work with more general forms of metrics (\ref{metric-schw-p}) and (\ref{metric-schw-n}) which in particular cases reduce to our mimetic solutions (\ref{metric-p}) and (\ref{metric-n}). They are also 
applicable for the case of exterior and interior Schwarzschild solutions. The reason is that our solutions (\ref{metric-p}) and (\ref{metric-n}) are different from the Schwarzschild solution through $A_{\pm}$ while the null hypersurfaces for metrics (\ref{metric-schw-p}) and  (\ref{metric-schw-n}) are determined by $B_{\pm}$.

We have two different Lorentzian manifolds $({\cal M}_{\pm},g_{\pm})$ and in local coordinates $x^\mu=(t_{\pm},r_{\pm},\theta,\varphi)$, the corresponding line elements are given by metrics (\ref{metric-schw-p}) and (\ref{metric-schw-n}). The surfaces $\Sigma_{\pm}\subset{\cal M}$ defined by $r_+ = 2m_+$ for (\ref{metric-schw-p}) and $t_- = 2 m_-$ for (\ref{metric-schw-n}) are determined by $B_{\pm}=0$. The normal vectors $K_{\pm \mu}$ to null hypersurfaces $\Sigma_{\pm}$ are given by
\begin{equation}\label{k}
K_{+\mu}=\partial_\mu r_+=\delta^{r_+}_\mu\,, \hspace{1cm}
K_{-\mu}=\partial_\mu t_-=\delta^{t_-}_\mu \,.
\end{equation}
From the above relations we can easily deduce that $\Sigma_{\pm}$ are null hypersurfaces since
\begin{equation}\label{k2-0}
K_{\pm\mu} K_{\pm}^\mu = B_\pm = 0\,,
\end{equation}
where $B_{\pm}$ are evaluated at the horizons $r_+ = 2m_+$ and $t_- = 2 m_-$ for $\Sigma_{\pm}$ respectively.

Our aim is to glue the manifolds $({\cal M}_+,g_+)$ and $({\cal M}_-,g_-)$ on null hypersurfaces $\Sigma_{\pm}$ where  the junction conditions have to be imposed.  The null hypersurfaces $\Sigma_{\pm}$ are determined by the parametric equations $x^{\mu}_{\pm}(y^a)$ where we implement the coordinates $y^a=(s,\theta^A)$ which are applicable on both $\Sigma_{\pm}$. Here $s$ is an arbitrary parameter which is not necessarily an affine parameter. The induced metrics $h_{\pm}$ on $\Sigma_{\pm}$ are the pullbacks of the embedding maps $\Sigma_{\pm}\to{\cal M}_{\pm}$ as
\begin{equation}\label{h-metrics}
h_{\pm{ab}} = g_{\pm\mu\nu}\, e^\mu_{\pm{a}} e^\nu_{\pm{b}} \,,
\end{equation}
where $e^\mu_{\pm{a}} \equiv \partial{x_{\pm}^{\mu}}/\partial{y^a}$ for null hypersurfaces with parametrization $dx_{\pm}^\mu = s K_{\pm}^\mu$ are, 
\begin{equation}\label{tangent-vectors}
e^\mu_{\pm{s}} = \Big(\frac{\partial{x_{\pm}^{\mu}}}{\partial{s}}\Big)\Big{|}_{\theta^A}
= K_{\pm}^\mu \,, \hspace{1cm}
e^\mu_{\pm{A}} = \Big(\frac{\partial{x_{\pm}^{\mu}}}{\partial{\theta^A}}\Big)\Big{|}_{s} \,,
\end{equation}
which satisfy the condition (\ref{k2-0}) and also $K_{\pm\mu}e^\mu_{\pm{A}} =0$. Using the above relations in Eq. (\ref{h-metrics}), we find that $h_{ss}=0=h_{sA}$. We therefore deal with a two-dimensional induced metric
\begin{equation}\label{sigma-metrics}
\sigma_{\pm{AB}} = g_{\pm\mu\nu}\, e^\mu_{\pm{A}} e^\nu_{\pm{B}}\,.
\end{equation}

The first junction condition for the null hypersurfaces is
\begin{equation}\label{first-JC}
\sigma_{+{AB}}|_{\Sigma_+} = \sigma_{-{AB}}|_{\Sigma_-}  \,.
\end{equation}
Let us check that whether the above junction condition can be satisfied for our solutions (\ref{metric-schw-p}) and (\ref{metric-schw-n}). In our local coordinates, the tangent vectors are given by
\begin{equation}
e^\mu_{+s} = B_+ \delta^\mu_{r_+} \,, \hspace{1cm} e^\mu_{-s} = B_- \delta^\mu_{t_-} \,, \hspace{1cm}
e^\mu_{\theta} = \delta^\mu_{\theta} \,, \hspace{1cm} e^\mu_{\varphi} = \delta^\mu_{\varphi} \,.
\end{equation}
Substituting from the above relation and also Eqs. (\ref{metric-schw-p}) and (\ref{metric-schw-n}) into the first junction condition (\ref{first-JC}), it is straightforward to show that the junction condition is satisfied through the identification $r_+\leftrightarrow{t_-}$ which  was expected from the beginning.

In order to check the second junction condition, we need to find the transverse curvature
\begin{equation}\label{transverse-C}
C_{\pm ab} \equiv \frac{1}{2} ( N_{\pm\mu;\nu} + N_{\pm\nu;\mu} ) e^\mu_{\pm{a}} e^\nu_{\pm{b}} \,,
\end{equation}
where the null vectors $N_{\pm}^\mu$ are defined as
\begin{equation}\label{N2-0}
N_{\pm\mu} N_{\pm}^\mu = 0\,, \hspace{1cm} N_{\pm\mu} K_{\pm}^\mu = -1 \,.
\end{equation}

Substituting from Eqs. (\ref{metric-schw-p}), (\ref{metric-schw-n}), and (\ref{k}) in above relations, we find the following solutions
\begin{equation}\label{N}
N_+^\mu = \Big( \mp \frac{1}{\sqrt{A_+ B_+ }} , -1 , 0 , 0 \Big) \,, \hspace{1cm}
N_-^\mu = \Big( -1 , \mp \frac{1}{\sqrt{A_- B_- }} , 0 , 0 \Big) \,.
\end{equation}
Correspondingly, substituting from Eqs. (\ref{tangent-vectors}) in (\ref{transverse-C}), we find the following nonzero components for the transverse curvature
\begin{eqnarray}\label{C}
C_{+11} = \frac{B'_+}{2} \,, \hspace{1cm} C_{+22} = - r_+ \,,
\hspace{1cm} C_{+33} = - r_+ \sin^2\theta \,, \nonumber \\
C_{-11} = - \frac{\dot{B}_-}{2} \,, \hspace{1cm} C_{-22} = - t_- \,,
\hspace{1cm} C_{-33} = - t_- \sin^2\theta \,.
\end{eqnarray}

The second junction condition is given by
\begin{equation}\label{second-JC}
C_{+ab} = C_{-ab} \,.
\end{equation}
It is straightforward to show that applying $r_+\leftrightarrow{t_-}$,   the above junction condition satisfies by the exterior and interior metrics  (\ref{metric-schw-p}) and (\ref{metric-schw-n}) at horizons $r_+ = 2m_+$ and $t_- = 2 m_-$ if we assume
\begin{equation}\label{mass-equivalence}
m_+= m_-\,.
\end{equation}

In summary, we have shown that the exterior solution (\ref{metric-schw-p}) and the interior solution (\ref{metric-schw-n}) are matched smoothly on the null hypersurfaces $r_+ = 2m_+$ and $t_- = 2 m_-$ where the junction conditions (\ref{first-JC}) and (\ref{second-JC}) are satisfied. We, therefore, have a black hole solution $({\cal M},g,\Sigma)$ where ${\cal M} = {\cal M}_{+} \cup {\cal M}_{-}$, $g=g_+\cup{g_-}$, and $\Sigma=\Sigma_{\pm}$ is the associated horizon. In addition to the metric, we have mimetic scalar field $\phi$ as dynamical variable in this scenario. Therefore, much similar to the junction conditions for the metric, we need to consider the appropriate junction condition for the scalar field as well. The scalar field satisfies first-order differential equation (\ref{mimetic-const.}) and we only need to check continuity of the scalar field. Integrating (\ref{phiprime-n}) we find solution for the scalar field while the solution for the exterior is given by (\ref{phi}). Equaling these solutions on the horizon $r=2m$, we find that they matchs each other for $c_3=2m$.

\begin{figure}
	\begin{center}
		\includegraphics[scale=0.45]{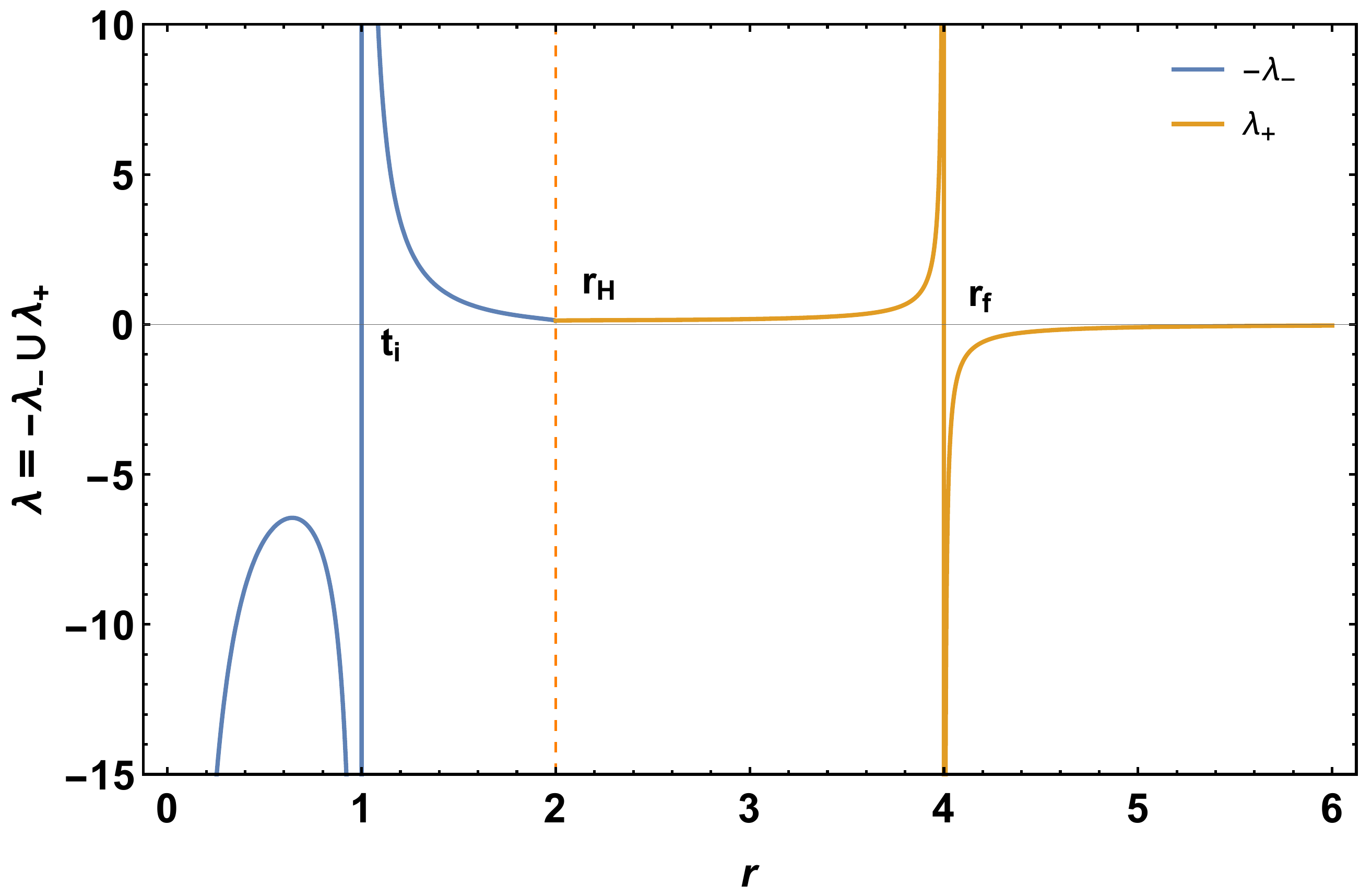}
		\caption{\label{fig2} The plot of $\lambda=-\lambda_{-}\cup\lambda_{+}$  for $m=1$, $t_i=1$ and $r_f=4$. We see that $\lambda$ is continuous across the surface of horizon at $r=2$. We deal with the black hole region $[t_i,2m]\cup[2m,r_f]$.}
	\end{center}	
\end{figure}

After satisfying the junction conditions, we expect that all curvature invariants to be continuous across the surface of horizon. However, it is instructive to check this explicitly. This is easy to verify  in our setup  since all curvature invariants (\ref{ricci-scalar}), (\ref{ricci-tensor-2}), (\ref{kretschmann}) and (\ref{weyl2}) are characterized by the auxiliary field $\lambda$ so we only need to check the continuity of $\lambda$ on the surface of horizon. For the exterior solution (\ref{metric-p}) all curvature invariants are the same as those obtained in Eqs. (\ref{ricci-scalar})-(\ref{weyl2}) via replacing $\lambda=\lambda_+$ where $\lambda_+$ is given by (\ref{lambda-rf-p}). For instance, for the Ricci scalar we have
\begin{equation}\label{Ricci-lambda-plus}
R_{+} = 2 \lambda_{+} \,.
\end{equation}
In the case of the interior solution (\ref{metric-n}), it is straightforward to check that all curvature invariants in the left hand side of (\ref{ricci-scalar}), (\ref{ricci-tensor-2}), (\ref{kretschmann}) and (\ref{weyl2}) have the same form as the right hand side but with $\lambda\to-\lambda_-$ where $\lambda_-$ is given by Eq. (\ref{lambda-tf}). We therefore will not write all of them explicitly here. Note that we have identified $m_+=m_-=m$. In the case of Ricci scalar, we can find it either by directly calculating it from Eq. (\ref{metric-n}) or replacing $\lambda\to-\lambda_-$ in Eq. (\ref{ricci-scalar}) as follows
\begin{equation}\label{Ricci-lambda-minus}
R_{-} = - 2 \lambda_{-} \,.
\end{equation}
We therefore have $R_{+}|_{\Sigma} = R_{-}|_{\Sigma}$ which from Eqs. (\ref{Ricci-lambda-plus}) and (\ref{Ricci-lambda-minus}) is equivalent to
\begin{equation}\label{lambda-JC}
\lambda_+|_{\Sigma} = - \lambda_-|_{\Sigma} \,.
\end{equation}
It is easy to explicitly confirm the above relation by evaluating $\lambda_+$ in Eq. (\ref{lambda-rf-p}) and $\lambda_-$ in Eq. (\ref{lambda-tf}) on the surface $r=2m$ which shows the continuity of the auxiliary field. Therefore, all curvature invariants are continuous across the surface of the horizon. The above relation holds irrespective of the values of $t_i$ and $r_f$.
In Fig. \ref{fig2} we have plotted the combined auxiliary field $\lambda=-\lambda_- \cup \lambda_+$  which confirms that $-\lambda_{-}$ matches to $\lambda_{+}$ at the surface of the apparent horizon.

\section{Caustic Singularities in the Mimetic Black Hole}\label{singularities}

In the previous section, we have constructed the mimetic black hole solution by gluing two different solutions. In this section, we investigate the physical properties of this mimetic black hole.

Let us first summarize the results from the previous section. The exterior metric  has the following static spherically symmetric form
\begin{eqnarray}\label{BH-exterior}
ds^2 = - \frac{dt^2}{4 r^4 \lambda_+^2} + \frac{dr^2}{1-\frac{2m}{r}} + r^2 d\Omega^2 \,,
\end{eqnarray}
where $\lambda_+$ is given by Eq. (\ref{lambda-rf-p}) replacing now $r_+$ with $r$ and $m_+$ with $m$. The interior solution is also given by
\begin{eqnarray}\label{BH-interior}
ds^2 = - \frac{d t^2}{\frac{2 m}{t}-1} + \frac{dr^2}{4 t^4 \lambda_-^2} + t^2 d\Omega^2 \,,
\end{eqnarray}
where $\lambda_-$ is given by Eq. (\ref{lambda-n}) replacing $t_-$ with $t$ and $m_-$ with $m$.

Now, we study the type of singularities at $t=t_i$ and $r=r_f$ that occur in our mimetic black hole solution (see Fig \ref{fig2}). Let us first study the singularity at $t=t_i$ for  the interior solution (\ref{BH-interior}).  We first note that the timelike vector $n^{\mu}=\partial^\mu\phi$ is orthogonal to the constant-time hypersurfaces and it is also normalized through the mimetic constraint Eq. (\ref{mimetic-const.}) as $n_{\mu}n^{\mu}=-1$. This vector then satisfies the geodesics equation
\begin{equation}\label{geodesics}
n^\mu \nabla_{\mu} n_{\nu} = 0 \,.
\end{equation} 

Since a congruence of geodesics  generally forms caustics, the above result shows that timelike hypersurfaces in mimetic setup with the action $S_{-}$  develop caustics. Our second aim is to show  when and where these caustic singularities  occur. In other words, from the above general result we cannot conclude that which of $t=0$ or $t=t_i$ is the time when caustics form. We therefore look at the metric (\ref{BH-interior}) noting  that at the time $t=t_i$, $g_{rr}$ vanishes while it does not vanish at $t=0$. Therefore, caustics with spherical symmetry are formed at finite time $t=t_i$  everywhere in background (see section 3 of Ref. \cite{mukohyama-caustics} for the similar caustic singularities that form in Minkowski spacetime with planar symmetry). Moreover, the second derivative (and higher derivative) of the field are characterized by the curvature invariants and the behaviour of all curvature invariants are given by $\lambda$. From Fig \ref{fig2}, we see that $\lambda$ and therefore all curvature invariants diverge(s) at the caustic points. Note that curvature invariants (or equivalently $\lambda$) also diverge at the point of standard singularity $t=0$ but $g_{rr}$ does not vanish there which shows that $t=0$ is not a caustic singularity. This is consistent with the intuitive discussions at the end of Section \ref{ext}.

A similar arguments can be applied to the exterior solution  (\ref{BH-exterior}). The vector $\partial^\mu\phi$ is orthogonal to constant-$r$ (timelike) hypersurfaces and it is normalized through the mimetic constraint Eq. (\ref{mimetic-const.}) as $n_{\mu}n^{\mu}=1$. Again we conclude that $n^\mu$ satisfies the geodesic equation (\ref{geodesics}) and we would have caustic singularities. To see when and where they  occur, we look at the exterior metric (\ref{BH-exterior}) where $g_{tt}$ vanish at $r=r_f$. We therefore conclude that, in contrast with the interior case, there are caustic singularities within {\it finite radius} of $r_f$ for all time in the background. The caustic singularities are localized at a finite radius for the exterior solution while they occur everywhere for the interior solution.

The general proof based on the geodesic equation (\ref{geodesics}) shows that caustic singularities are inevitable in both mimetic solutions defined by the actions $S_{\pm}$ in (\ref{action}). Although we have studied the details of caustic singularities in the case of spherical symmetric and anisotropic time-dependent backgrounds, the above proof shows that the existence of these singularities is  independent of the background. The Appearance of caustics is not pathological for an effective field theory while it is a serious problem for a fundamental theory. Much similar to the metric field, the scalar field $\phi$ is a fundamental  field in mimetic scenario which can couple to Standard Model fields. The mimetic scenario then breaks down near the caustics singularities. This problem can be solved if we suppose that the mimetic theory is an effective field theory which only emerges at low energies in the IR regime. Then, the appearance of the caustics signals for the emergence of a  new physics at ultraviolet (UV) regime. In this regards, there would be a UV completion theory for the mimetic scenario which governs high curvature regime and prevents the formation of caustics. In the case of $S_{-}$, indeed, Horava-Lifshitz gravity can be thought as a UV completion theory \cite{Ramazanov:2016xhp}. Interestingly, caustics would not form at high curvature regimes in the Horava-Lifshitz gravity scenario (see Ref. \cite{mukohyama-caustics} for the details). A similar UV completion theory, but with different foliations (in comparison with the spcaelike foliation in Horava-Lifshitz gravity) which respects the symmetries of $S_{+}$ can potentially prevent the formation of caustics in the exterior solution.  In this paper, however, we treat mimetic gravity as an effective theory and restrict ourselves to the low curvature regime of $[t'_i,2m]\cup[2m,r'_f]$ with $t'_i \gg t_i$ and $r'_f \ll r_f$ so $t'_f$ and $r'_f$ are sufficiently far from the points of caustic singularities $t_i$ and $r_f$. Moreover, as we have already mentioned, the intervals $0<t<t_i$ and $r_f<r<\infty$ have negative curvatures and therefore they are physically different than the region $[t_i,2m]\cup[2m,r_f]$ which posses positive curvature. 

To get some physical intuition, it is useful to model the effects of mimetic field by a fluid. Let us start with the interior solution which looks like an anisotropic cosmological model. The right hand side of the Einstein equations (\ref{Einstein-Eqs}) can be viewed as an effective energy-momentum tensor which has the same covariant form for both  $S_{+}$ and $S_{-}$ actions. Substituting from Eqs. (\ref{phiprime-n}) and (\ref{metric-n}), we find that for the interior solution $T^\mu_\nu=\mbox{diag}(2\lambda,0,0,0)$. This shows that the energy density is given by $\rho=-2\lambda$ while the pressure vanishes. A vanishing pressure in region $t_i<t<2m$ is the signature of attractive gravity which usually leads to caustic singularity within finite time as we have already mentioned. Since $\lambda>0$ for $0<t<t_i$, this region corresponds to negative energy density $\rho<0$ which is not allowed physically. Similarly, the region $r_f<r<\infty$ is the exterior counterpart of the interior region $0<t<t_i$. However, we cannot model it by a known matter field. More precisely, substituting from Eqs. (\ref{phiprime}) and (\ref{metric-p}), we find that for the exterior region $T^\mu_\nu=\mbox{diag}(0,-2\lambda,0,0)$, so it does not take the form of a perfect fluid as the pressure is not the same in the radial and angular directions. Moreover, in the region $r_f<r<\infty$, the energy density component vanishes. Although this is not pathological but we do not know any realistic matter field (fluid) which represents this kind of property.

To understand thermodynamical properties of this solution, we look at the Killing vectors of the exterior and interior solutions. For the exterior solution (\ref{BH-exterior}), there are four Killing vectors among which three of them are associated with the spherical symmetry and one  corresponds to the static nature of the metric with the Killing vector ${\cal R}^\mu=\delta^\mu_t$ which we are interested in. Similarly, for the interior solution (\ref{BH-interior}), ${\cal T}^\mu=\delta^\mu_r$ is a Killing vector that corresponds to the fact that the solution is $r$-independent. From Eqs. (\ref{BH-exterior}) and (\ref{BH-interior}), we then find that
\begin{eqnarray}\label{Killing-vector-ext}
{\cal R}_\mu {\cal R}^\mu &=& g_{tt} = - \frac{1}{4r^2\lambda_+^2}\, , \\
\label{Killing-vector-int}
{\cal T}_\mu {\cal T}^\mu &=& g_{rr} = \frac{1}{4t^2\lambda_-^2} \,.
\end{eqnarray}

The Killing horizon is defined as the location of the  points where the magnitude of the Killing vectors vanish. From Eqs. (\ref{lambda-rf-p}) and (\ref{lambda-tf}) we see that the above Killing vectors vanish only asymptotically as
\begin{eqnarray}\label{Killing-vector-ext-vanish}
{\cal R}_\mu {\cal R}^\mu |_{r=r_f} &=& - \frac{1}{4r_f^2\lambda_+^2(r_f)} \to 0 \,, \\
\label{Killing-vector-int-vanish}
{\cal T}_\mu {\cal T}^\mu |_{t=t_i} &=& \frac{1}{4t_i^2\lambda_-^2(t_i)} \to 0 \,.
\end{eqnarray}
We see that at the surfaces of caustic singularities, $r=r_f$ and $t=t_i$, Killing vectors become null since $|\lambda_{\pm}|\to\infty$. However, we cannot trust the mimetic theory near the caustics and to study the thermodynamic properties of this black hole solution one needs a UV completion theory.

\section{Summary and Conclusions}

Mimetic scenario is a conformal extension of general relativity in which the longitudinal mode of gravity, encoded in the mimetic scalar field $\phi$, becomes dynamical even in the absence of any matter field. This extra degree of freedom plays the roles of dark matter in a cosmological background. 

In this paper, we have focused on the static spherically symmetric background to see whether the mimetic setup can admit a black hole solution. We found that there is not a static black hole solution as far as mimetic scalar field contributes to the background. This can be intuitively understood if we note that mimetic matter plays the role of dark matter in cosmological background which forbids a static configuration. Thus, assuming that mimetic scalar field does not contribute to the background, we found a stealth Schwarzschild solution with non-trivial scalar field profile. On the other hand, we have shown that it is also possible to find a static solution even if mimetic scalar field contributes to the background. The solution, however, is not a black hole but a spacetime with naked singularities. We have shown that one of these singularities is similar to the singularity at the center of standard Schwarzschild black hole while the other is a caustic type singularity. The mimetic theory then breaks down near the caustic singularity. We have concluded that there is no natural black hole solution in this scenario if mimetic scalar field contributes to the background. To resolve this problem, we then considered two types of mimetic models with the actions $S_{+}$ and $S_{-}$ in which the vector $\partial^\mu\phi$ is spacelike in $S_{+}$ while it is timelike in $S_{-}$. We have shown that the first setup admits a static spherically symmetric solution with a horizon while the latter admits a time-dependent homogeneous anisotropic solution with a horizon. We have verified that these two different solutions (metrics (\ref{BH-exterior}) and (\ref{BH-interior}) for the exterior and interior respectively)  can be glued at their joint horizon leading to a mimetic black hole solution. Both of the exterior and interior solutions suffer from caustic singularities at high curvature regime. The mimetic scenario then would be considered as an effective theory which is applicable in the IR regime and breaks down near the caustic singularities. One then needs a UV-completion theory for the mimetic scenario near the caustic singularities. In the case of interior solution, the Horava-Lifshitz theory can be considered as a UV-completion of the timelike mimetic model $S_{-}$. Inspired by this fact, for the spacelike mimetic model $S_{+}$, we speculated that a UV-completion theory similar to the the Horava-Lifshitz but with different foliation can be constructed. However, we do not know such a scenario.

There are two different Killing horizons for this black hole solution in which non of them coincides with the apparent horizon. Based on our intuition from general relativity, there would be no notion of Hawking radiation to be associated with the apparent horizon and one needs to look for Hawking radiation at Killing horizons. However, the intuition from general relativity may not be applicable in our scenario. Moreover, these Killing horizons turned out to be singular hypersurfaces due to caustics formation which signal for a new physics. Therefore, we may not trust the theory around these singular hypersurfaces.

\vspace{1cm}
{\bf Acknowledgments:} We would like to thank Shinji Mukohyama for correspondence and comments on the draft and Seyed Ali Hosseini Mansoori for useful discussions. The work of M. A. G. was supported by Japan Society for the Promotion of Science (JSPS) Grants-in-Aid for international research fellow No. 19F19313.

\vspace{0.7cm}
\appendix

\section{Mimetic Gravity as Singular Limit of Conformal Transformation}\label{app-a}
\setcounter{equation}{0}
\renewcommand{\theequation}{A\arabic{equation}}

In this appendix we show that tranformations (\ref{mimetic-trans}) can be uinquely obtained from the singular limit of the general conformal transformation 
\begin{equation}\label{conformal-trans}
g_{\mu\nu} = A(\phi,X) \, {\tilde g}_{\mu\nu} \,,
\end{equation}
where $X={\tilde g}^{\alpha\beta}\partial_{\alpha}\phi \partial_\beta\phi$ and the inverse metric is $g^{\mu\nu} = A^{-1} \, {\tilde g}^{\mu\nu}$. The original mimetic scenario is shift symmetric and, indeed, the appearance of the dark matter-like energy density component is the consequence of shift symmetry \cite{Mirzagholi:2014ifa}. We, however, consider the most general case and only at the end concentrate on the shift-symmetric subset.

To find the singular limit, we need to look at the Jacobian of the transformation $\frac{\partial g_{\mu\nu}}{\partial {\tilde g}_{\alpha\beta}}$ to see whether it vanishes. Therefore, we look at the eigenvalue equation for the determinant of the Jacobian which is given by \cite{Zumalacarregui:2013pma}
\begin{equation}\label{eigen}
\left(\frac{\partial g_{\mu\nu}}{\partial {\tilde g}_{\alpha\beta}}
- e^{(n)}\, \delta_\mu^\alpha\delta_\nu^\beta\right)\xi^{(n)}_{\alpha\beta}=0 \,,
\end{equation}
where $e^{(n)}$ are the eigenvalues and $\xi^{(n)}_{\mu\nu}$ are the corresponding eigentensors. Substituting Eq. (\ref{conformal-trans}) in the above eigenvalue equation, we find
\begin{equation}\label{eigen-c}
(A-e^{(n)})\xi_{\mu\nu}^{(n)} - \big( A_{,X}
\xi_{\alpha\beta}^{(n)} \partial^\alpha\phi \partial^\beta\phi \big)  g_{\mu\nu} = 0 \,.
\end{equation}

The above equation has ten solutions for eigenvectors $\xi_{\alpha\beta}^{(n)}$ which can be classified into the two different types of conformal and kinetic solutions \cite{Zumalacarregui:2013pma}. The non-trivial solution, which is relevant for the mimetic scenario, is given by \cite{Firouzjahi:2018xob}
\begin{equation}\label{eigenvalue-Kphipsi}
e = A - X A_{,X} \,, \hspace{1cm} \mbox{with}
\hspace{1cm} \xi_{\mu\nu} = {\tilde g}_{\mu\nu} \,.
\end{equation}

The singular limit $e = 0$ then gives the following solution,
\begin{equation}\label{singular-solution}
A = p(\phi) \, X \,.
\end{equation}

Since the above solution is linear in $X$, the physical metric $g_{\mu\nu}$ is invariant under conformal transformation of the auxiliary metric ${\tilde g}_{\mu\nu}$. Substituting the above result in (\ref{conformal-trans}), we find 
\begin{equation}\label{conformal-trans-sol}
g_{\mu\nu} = p(\phi) \, \big( {\tilde g}^{\alpha\beta}\partial_{\alpha}\phi\partial_{\beta}\phi \big) \, {\tilde g}_{\mu\nu} \,,
\end{equation}
which implies the following constraint for the physical metric
\begin{equation}\label{mimetic-constraint-general}
p(\phi) \, \big( g^{\alpha\beta}\partial_{\alpha}\phi\partial_{\beta}\phi \big) = 1 \,.
\end{equation}

Transformation (\ref{conformal-trans-sol}) or equivalently constraint (\ref{mimetic-constraint-general}) can be thought as a generalization of the mimetic gravity so that we recover the original mimetic scenario \cite{Chamseddine:2013kea} in the particular case of $p(\phi)=-1$. In general, the functional form of $p(\phi)$ cannot be determined by the principle of conformal invariance. Assuming shift-symmetry then $p= \mbox{constant}$, but the transformation (\ref{conformal-trans-sol}) fixes up to the sign of $p$ since the constant $p$ can be removed through the redefinition of the scalar field. Relations (\ref{conformal-trans-sol}) and (\ref{mimetic-constraint-general}) then reduce to (\ref{mimetic-trans}) and (\ref{mimetic-const.}) respectively.

\section{Anisotropic Mimetic Cosmologies}\label{appendix}
\setcounter{equation}{0}
\renewcommand{\theequation}{B\arabic{equation}}

\begin{figure}
	\centering
	\includegraphics[width=300 pt]{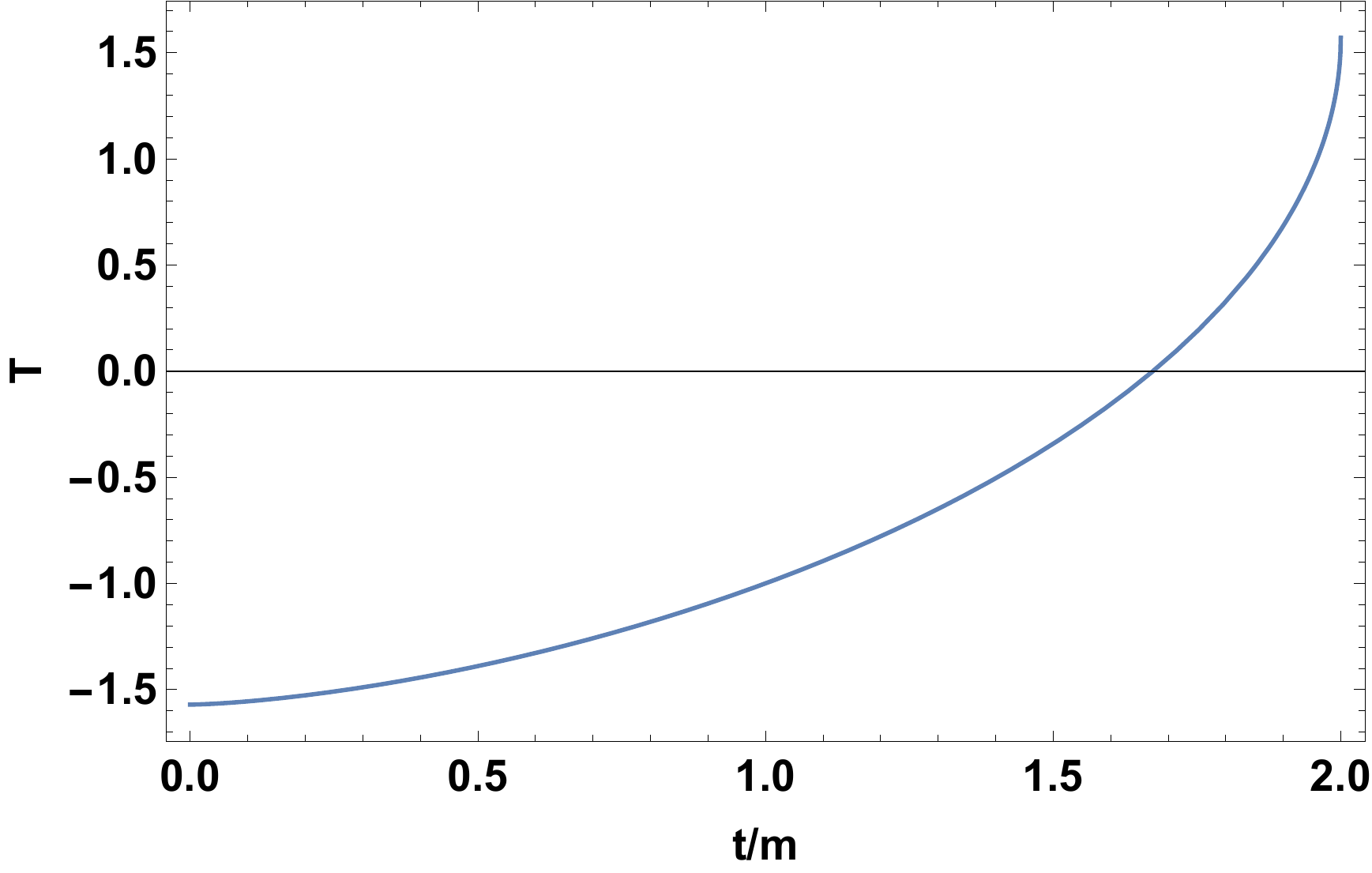}
	\caption{The new time coordinates $T$ as a function of the old time coordinate $t/m$. We identify $T=0$ with $t=t_i$ so that our solutions here with $T\in[0,+\frac{m\pi}{2}]$ covers $t\in[t_i,2m]$ in Fig.~\ref{fig2}, corresponding to the interior of the black hole. }
	\label{T-fig2}
\end{figure}

In order to obtain an interior solution for our black hole, we have considered mimetic model $S_{-}$ with homogeneous {\it anisotropic} metric ansatz  (\ref{metric-schw-n}) in local coordinates $(t_-,r_-,\theta,\varphi)$
\begin{align}\label{inte}
ds^2 = -\frac{dt^2}{B_{-}(t)}+A_{-}(t)dr^2+t^2 d\Omega^2\,,
\end{align}
where we have identified $(t_{-},r_{-})$ with $(t,r)$ for the sake of simplicity which is consistent with the notation that we have used in Section \ref{singularities}. The local coordinates used in (\ref{inte}) was more apt in gluing the  exterior (\ref{metric-p}) and interior (\ref{metric-n}) solutions. On the other hand, we know that the homogeneous {\it isotropic} cosmological solution for the mimetic model $S_{-}$ is the mimetic dark matter in a FLRW background \cite{Chamseddine:2013kea}. In order to compare our anisotropic results with its isotropic counterpart, it is better to work in another coordinates. In this respect, we consider the following Kantowski-Sachs spacetime 
\begin{align}\label{k-s}
ds^2=-dT^2+X(T)^2dr^2+Y(T)^2 d\Omega^2\,, 
\end{align}
where the new time coordinate is defined as
\begin{align}\label{trans}
dT = \frac{dt}{\sqrt{B_{-}}}\,,
\end{align}
and $X(T)\equiv\sqrt{A_{-}(T)}$ and $Y(T)\equiv t(T)$ are two scale factors. To see the relation between the new time coordinate $T$ and the old time coordinate $t$, we substitute from (\ref{B-n}) and then integrate (\ref{trans}) to obtain
\begin{align}\label{T-t}
T = - \sqrt{t} \sqrt{2 m-t}-m\, \arctan\left(-\frac{t-m}{\sqrt{t}\sqrt{2m-t}}\right)\,,
\end{align}
where we have set a constant of integration to zero which corresponds to a shift in $T$. From the above solution, we see that $T\in[-\frac{m\pi}{2},+\frac{m\pi}{2}]$ for $t\in[0,2m]$. In Fig. (\ref{T-fig2}), we have plotted $T$ as a function of  $t/m$. From Eq. (\ref{lambda-tf}), we know that $t=t_i$ is an arbitrary constant which determines the time that separates two regions with different curvatures. We have discussed that the region $t\in[0,t_i)$  violates the energy conditions if we model the mimetic field with a fluid with vanishing pressure. Therefore, on the physical ground, it is reasonable to identify $t=t_i$ with $T=0$ and restrict ourselves to the positive branch of $T$ so that  $T\in[0,+\frac{m\pi}{2}]$.

From Eqs. (\ref{phi-lambda-t}) and (\ref{trans}), we see that the consistent ansatz for the metric (\ref{k-s}) would be
\begin{equation}\label{phi-lambda-T}
\phi=\phi(T) \,, \hspace{1cm} \lambda_-=\lambda_-(T) \,.
\end{equation}

Substituting Eq. (\ref{phi-lambda-T}) together with Eq. (\ref{k-s}) in the mimetic constraint Eq. (\ref{mimetic-const.}) for the  $S_{-}$ model, we find $(d\phi/dT)^2=1$ that after integration yields 
\begin{align}\label{phi-T}
\phi(T)=T\,,
\end{align}
where we have chosen the plus sign in the above solution.

The equation of motion for the mimetic scalar field from Eq.~(\ref{KG-Eqs}) then yields
\begin{align}\label{x}
X(T)=\frac{C_1}{Y(T)^2\lambda_-(T)}\,,
\end{align}
where $C_1$ is a constant of integration.

The $TT$ and $rr$ components of the Einstein equations give
\begin{align}\label{lamb}
2 (XY)^{-1} \frac{dX}{dT}\frac{dY}{dT}+Y^{-2} \bigg( 1+\Big(\frac{dY}{dT}\Big)^2 \bigg) = - 2 \lambda_- \,,
\end{align}
and
\begin{align} \label{rr}
2 Y \frac{d^2Y}{dT^2} + \Big(\frac{dY}{dT}\Big)^2 +1 =0 \,.
\end{align} 

The solution to Eq. (\ref{rr}) is given in terms of the inverse function $f^{-1}(T)$ as follows
\begin{align}\label{y}
Y(T)=f^{-1}(T)\, ; \hspace{1cm}
f(T)=2C_{2}\arctan\Big(\frac{\sqrt{T}}{\sqrt{2C_{2}-T}} \Big) -\sqrt{2C_2T-T^2} \,,
\end{align}
where $C_2$ is another integration constant. Identifying the point $T=2C_2$ with $t=2m$, from (\ref{T-t}) we find that $C_2=m\pi/4$.

\begin{figure}[!ht]
	\includegraphics[scale=0.5]{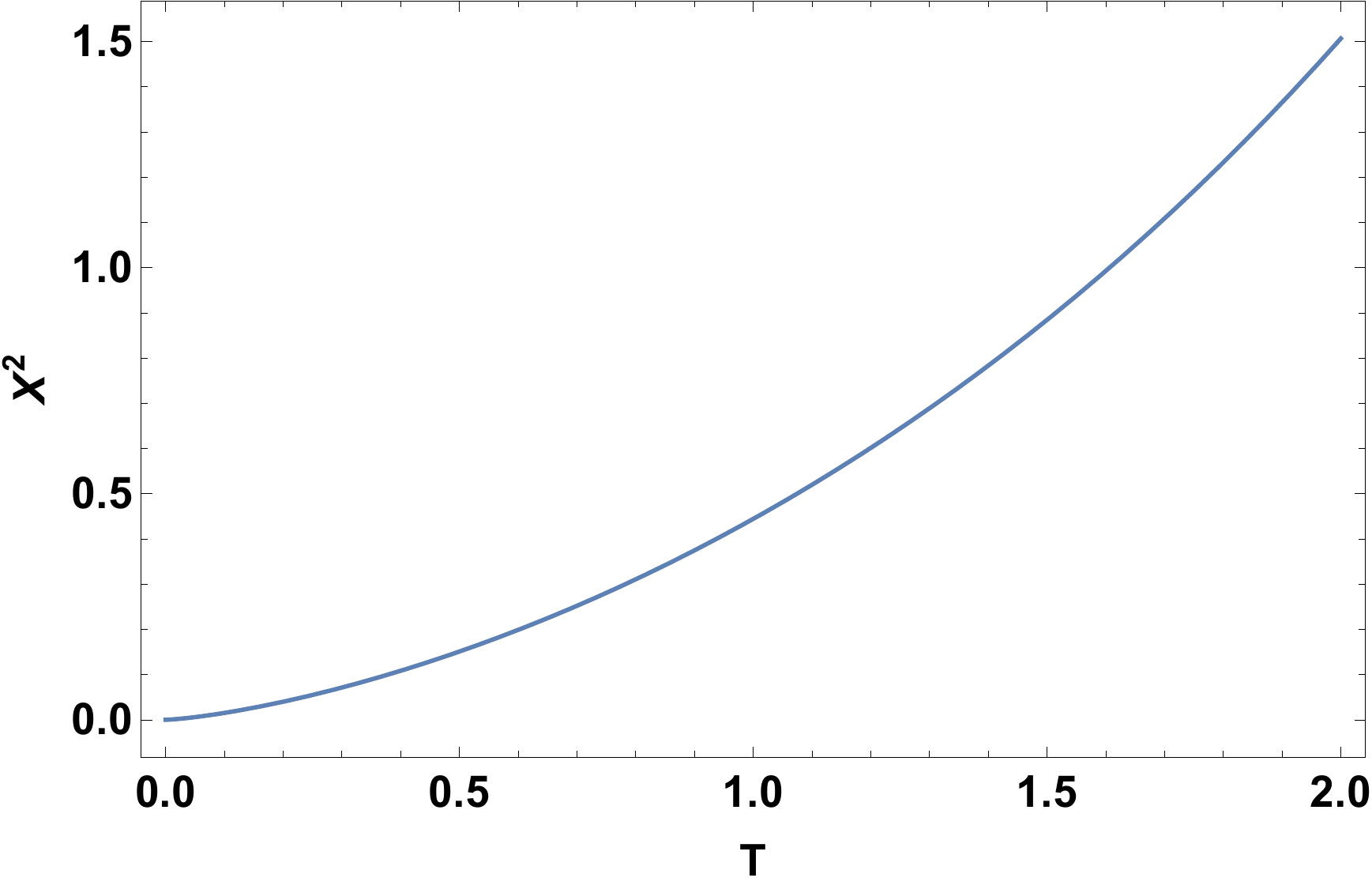}
	\includegraphics[scale=0.5]{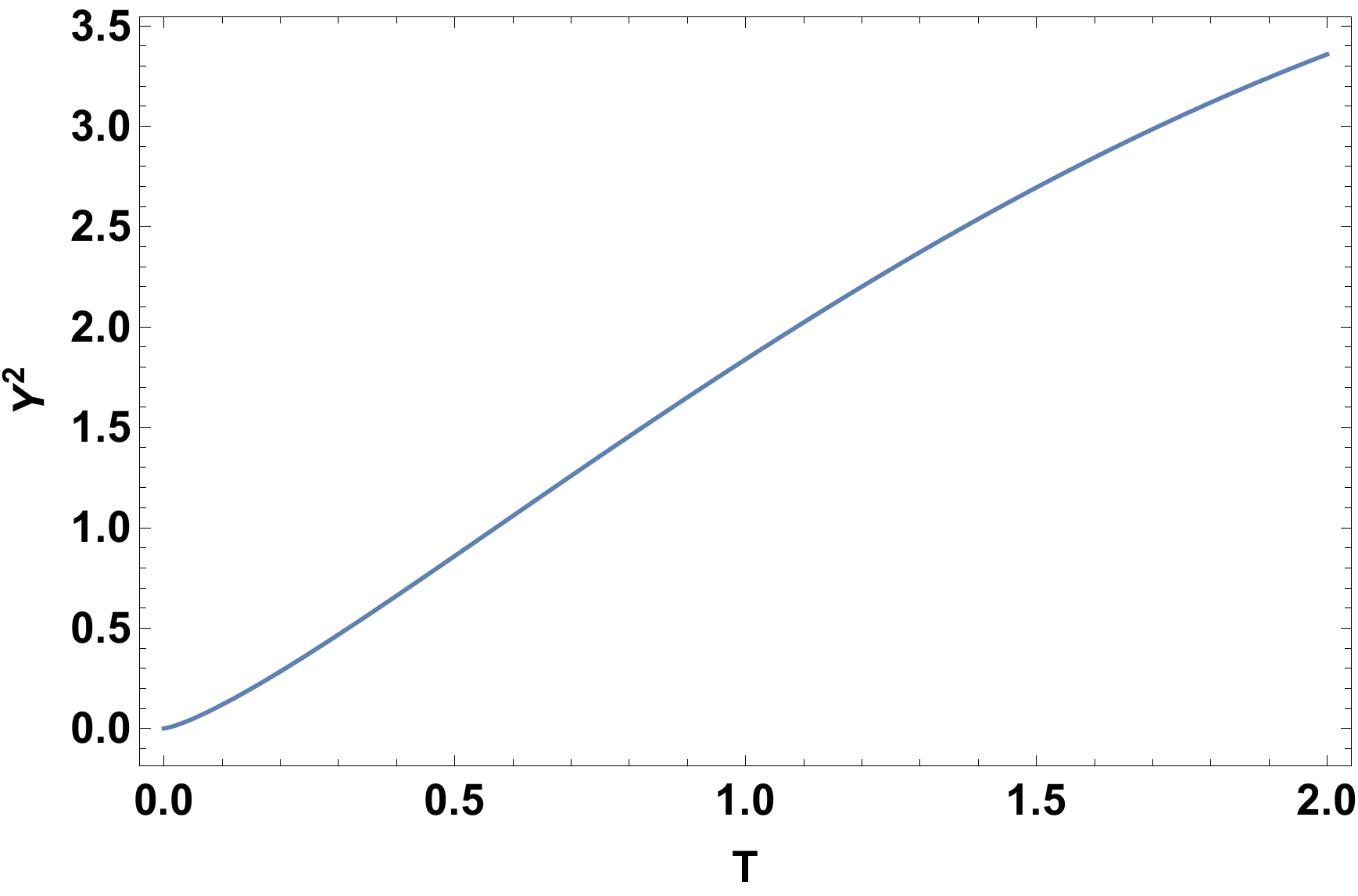}
	\includegraphics[scale=0.5]{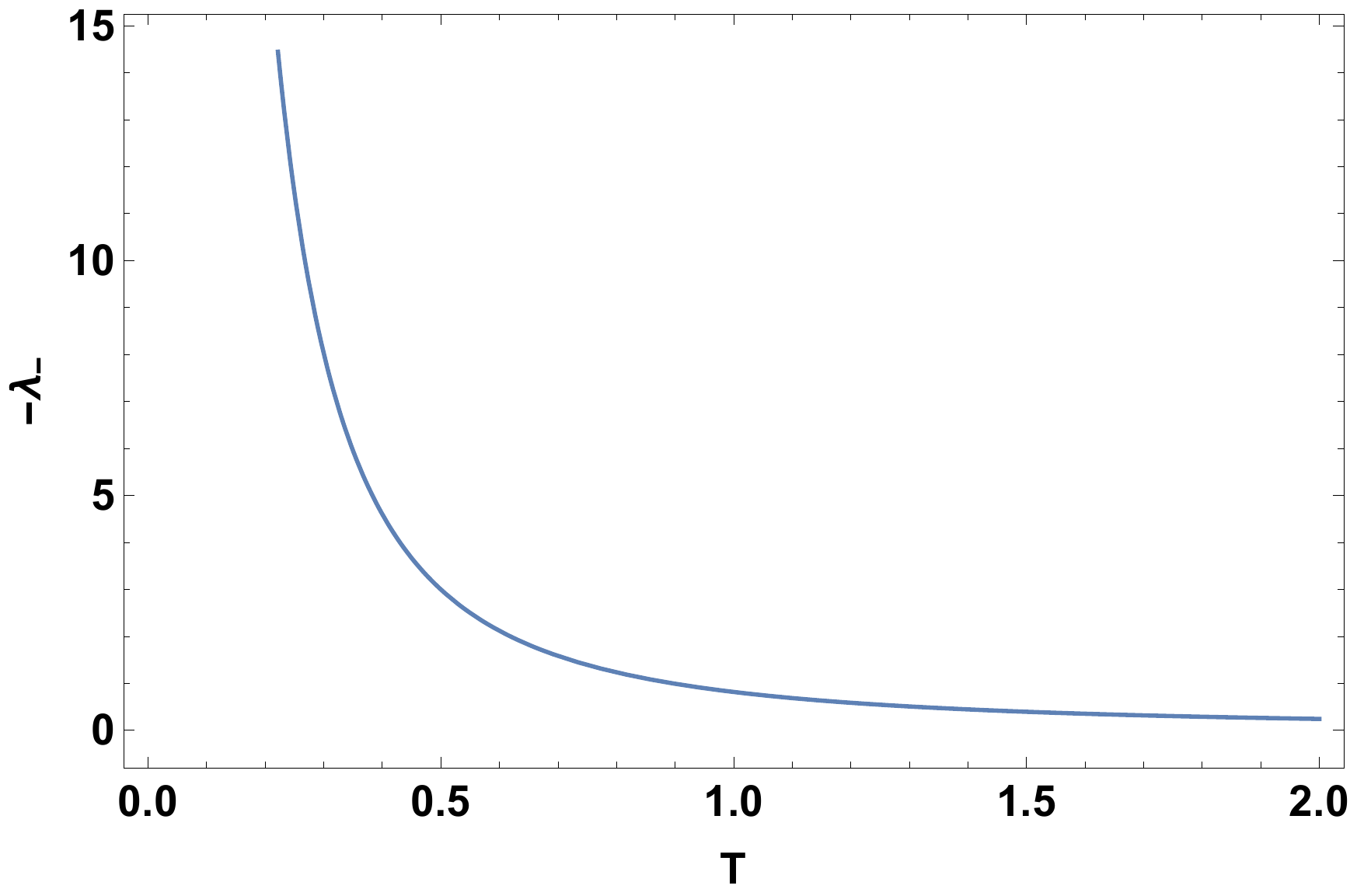}
	\includegraphics[scale=0.5]{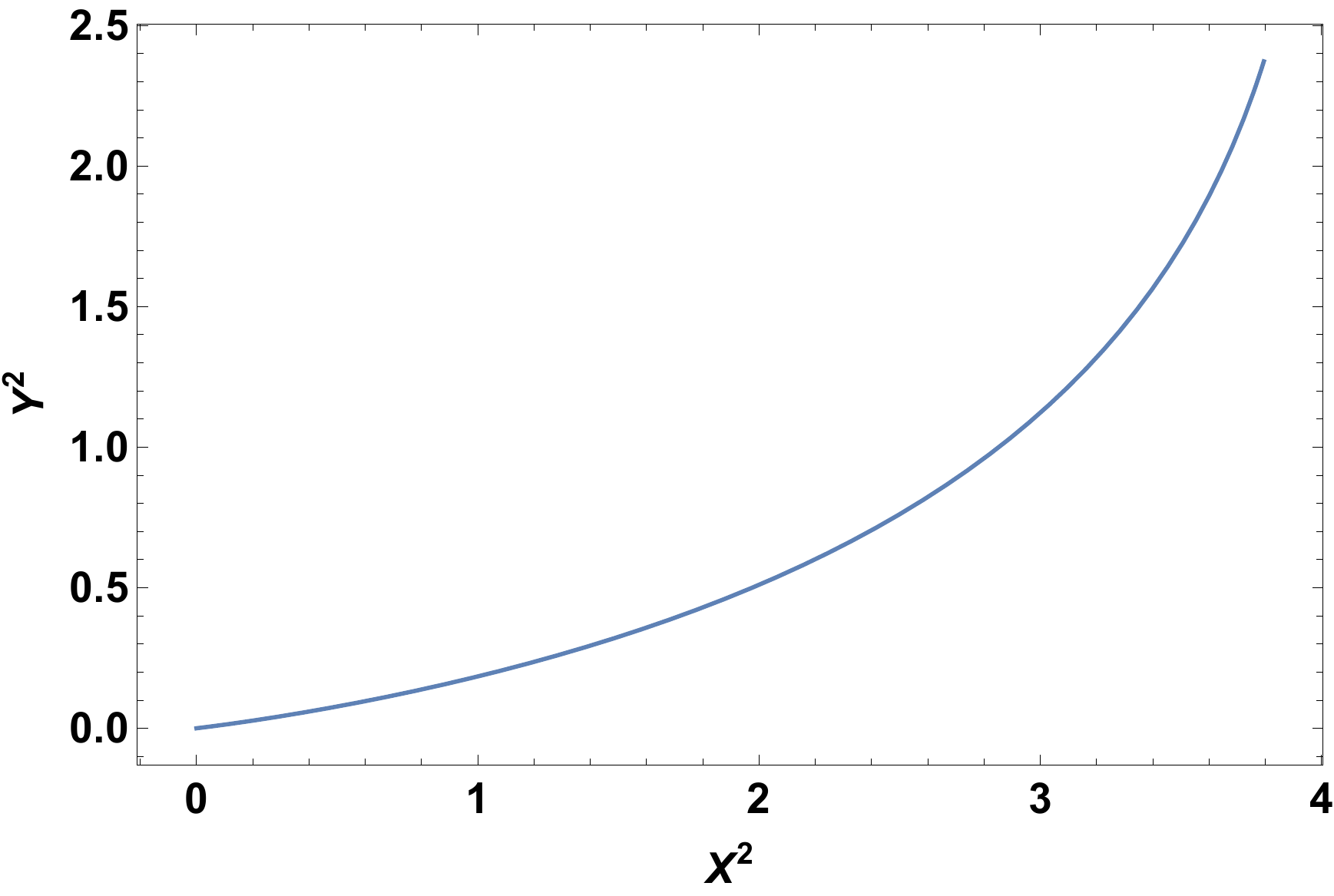}
	\caption{ The plots of $X^2$ ({top left}), $Y^2$ ({top right}), $-\lambda_{-}$ ({bottom left}) 
	as functions of $T$ and the plot of $X^2$ vs. $Y^2$ ({bottom right}).		
		As the scale factors shrink to zero, the energy density of the mimetic field diverges producing caustic singularities. $X^2$ and $Y^2$ both expand or contract simultaneously. The behavior of $-\lambda_-$ shows that the solution in this coordinate covers only $t_i<t$. The figures are plotted for $C_1=1=C_2$.}\label{fig}
\end{figure}

Substituting Eq. (\ref{y}) into Eqs. (\ref{x}) and (\ref{lamb}), we can solve for both $X$ and $\lambda_-$ as a function of $T$. Having obtained these solutions, our dynamical system is completely solved. We have plotted the solutions for $X^2$, $Y^2$, and $\lambda_-$ as functions of $T$ in Fig.~\ref{fig}. From Fig \ref{fig2} we see that at $T=0$, $X,Y\to0$ and $\lambda\to\infty$. These results can be directly confirmed from the equations (\ref{x})-(\ref{y}). From (\ref{y}), we see that $Y(T=0)=f^{-1}(T=0)=0$. Substituting (\ref{x}) and (\ref{y}) in (\ref{lambda}) and then solving for $X(T)$, it is straightforward to show that $X\propto{T}^{2/3}$ in the limit $T\to0$ and therefore $X(T=0)=0$. This singularity is nothing but the formation of caustics. The metric components $g_{rr}$, $g_{\theta\theta}$, and $g_{\varphi,\varphi}$ all vanish in metric (\ref{k-s}) at $T=0$ while all curvature invariants, which are proportional to the second derivative of metric, diverge since they are characterized by $\lambda$. Therefore caustics form at the finite time $T=0$ everywhere in the background.

Let us now compare the results of this appendix with those  obtained for  mimetic dark matter scenario in FLRW background \cite{Chamseddine:2013kea}. First of all, the metric (\ref{k-s}) becomes isotropic for $X\propto Y$ and then $X$ (or $Y$) plays the roles of the scale factor $a(T)$ in FLRW background. In this limit, from Eq. (\ref{x}) we find $\lambda_{-}\propto X^{-3}\propto a^{-3}$ which shows that the energy density $\rho=-2\lambda_{-}\propto{a}^{-3}$. This is nothing but the dark matter energy density. This can also be seen if we look at the metric (\ref{k-s}) in the limit of small $T$
\begin{align}\label{metric-small-T}
ds^2 \approx - dT^2 + \frac{T^{\frac{4}{3}}}{6^{\frac{2}{3}}C_2^{\frac{4}{3}}} \Big( dr^2 + 9C_2^2 d\Omega^2 \Big)\,.
\end{align}
The above metric shows that for small $T$, the spacetime in the vicinity of the singularity $T=0$ (which corresponds to $t=t_i$ in Fig.~\ref{fig2}) behaves like a matter dominated universe with the scale factor $a(T)\sim T^{\frac{2}{3}}$, as expected.

Moreover, we note that due to the presence of a dark matter-like fluid, the metric (\ref{metric-small-T}) is contracting in both  $r$ and  $(\theta,\varphi)$ directions. This is in contrast to the Schwarzschild solution for the interior of the black hole which is expanding along the $r$ direction but contracting in $(\theta,\varphi)$ directions. It also signals the formation of caustic singularities due to the attractive nature of dark matter-like fluid.

{}


\begin{thebibliography}{}


\bibitem{Chamseddine:2013kea}
  A.~H.~Chamseddine and V.~Mukhanov,
  JHEP {\bf 1311}, 135 (2013),
  [arXiv:1308.5410 [astro-ph.CO]].


\bibitem{Deruelle:2014zza}
N.~Deruelle and J.~Rua,
JCAP {\bf 1409}, 002 (2014), 
[arXiv:1407.0825 [gr-qc]].


\bibitem{Domenech:2015tca} 
G.~Domènech, S.~Mukohyama, R.~Namba, A.~Naruko, R.~Saitou and Y.~Watanabe,
Phys.\ Rev.\ D {\bf 92}, no. 8, 084027 (2015), 
[arXiv:1507.05390 [hep-th]].


\bibitem{Firouzjahi:2018xob} 
H.~Firouzjahi, M.~A.~Gorji, S.~A.~Hosseini Mansoori, A.~Karami and T.~Rostami,
JCAP {\bf 1811}, 046 (2018)
doi:10.1088/1475-7516/2018/11/046
[arXiv:1806.11472 [gr-qc]].

\bibitem{Shen:2019nyp}
L.~Shen, Y.~Zheng and M.~Li,
JCAP \textbf{12}, 026 (2019)
doi:10.1088/1475-7516/2019/12/026
[arXiv:1909.01248 [gr-qc]].


\bibitem{Golovnev:2013jxa}
A.~Golovnev,
Phys.\ Lett.\ B {\bf 728}, 39 (2014),
[arXiv:1310.2790 [gr-qc]].


\bibitem{mimetic}
F.~Arroja, N.~Bartolo, P.~Karmakar and S.~Matarrese,
JCAP {\bf 1509}, 051 (2015), 
[arXiv:1506.08575 [gr-qc]].

\bibitem{Sadeghnezhad:2017hmr} 
N.~Sadeghnezhad and K.~Nozari,
Phys.\ Lett.\ B {\bf 769}, 134 (2017),
[arXiv:1703.06269 [gr-qc]].

\bibitem{Liu:2017puc} 
D.~Langlois, H.~Liu, K.~Noui and E.~Wilson-Ewing,
Class.\ Quant.\ Grav.\  {\bf 34}, no. 22, 225004 (2017),
[arXiv:1703.10812 [gr-qc]].

\bibitem{Dutta:2017fjw} 
J.~Dutta, W.~Khyllep, E.~N.~Saridakis, N.~Tamanini and S.~Vagnozzi,
JCAP {\bf 1802}, 041 (2018),
[arXiv:1711.07290 [gr-qc]].

\bibitem{Casalino:2018tcd} 
A.~Casalino, M.~Rinaldi, L.~Sebastiani and S.~Vagnozzi,
Phys.\ Dark Univ.\  {\bf 22}, 108 (2018),
[arXiv:1803.02620 [gr-qc]].

\bibitem{Brahma:2018dwx} 
S.~Brahma, A.~Golovnev and D.~H.~Yeom,
Phys.\ Lett.\ B {\bf 782}, 280 (2018),
[arXiv:1803.03955 [gr-qc]].

\bibitem{deHaro:2018sqw} 
J.~de Haro, L.~Aresté Saló and S.~Pan,
Gen.\ Rel.\ Grav.\  {\bf 51}, no. 4, 49 (2019),
[arXiv:1803.09653 [gr-qc]].

\bibitem{deCesare:2018cts} 
M.~de Cesare,
Phys.\ Rev.\ D {\bf 99}, no. 6, 063505 (2019),
[arXiv:1812.06171 [gr-qc]].

\bibitem{Ganz:2019vre}
A.~Ganz, N.~Bartolo and S.~Matarrese,
JCAP \textbf{12}, 037 (2019)
doi:10.1088/1475-7516/2019/12/037
[arXiv:1907.10301 [gr-qc]].

\bibitem{Myrzakulov:2015qaa} 
R.~Myrzakulov, L.~Sebastiani and S.~Vagnozzi,
Eur.\ Phys.\ J.\ C {\bf 75}, 444 (2015)
doi:10.1140/epjc/s10052-015-3672-6
[arXiv:1504.07984 [gr-qc]].

\bibitem{Schutz:1970my} 
B.~F.~Schutz,
Phys.\ Rev.\ D {\bf 2}, 2762 (1970).
doi:10.1103/PhysRevD.2.2762

\bibitem{Brown:1992kc} 
J.~D.~Brown,
Class.\ Quant.\ Grav.\  {\bf 10}, 1579 (1993)
doi:10.1088/0264-9381/10/8/017
[gr-qc/9304026].

\bibitem{Barvinsky:2013mea} 
A.~O.~Barvinsky,
JCAP {\bf 1401}, 014 (2014)
doi:10.1088/1475-7516/2014/01/014
[arXiv:1311.3111 [hep-th]].

\bibitem{mukohyama-caustics} 
S.~Mukohyama,
JCAP {\bf 0909}, 005 (2009)
doi:10.1088/1475-7516/2009/09/005
[arXiv:0906.5069 [hep-th]].

\bibitem{caustics}
F.~Capela and S.~Ramazanov,
JCAP {\bf 1504}, 051 (2015), 
[arXiv:1412.2051 [astro-ph.CO]].

\bibitem{Vikman:2017gxs} 
A.~Vikman,
arXiv:1712.10311 [astro-ph.CO].

\bibitem{Gorji:2018okn} 
M.~A.~Gorji, S.~Mukohyama, H.~Firouzjahi and S.~A.~Hosseini Mansoori,
JCAP {\bf 1808}, no. 08, 047 (2018),
[arXiv:1807.06335 [hep-th]].

\bibitem{Jirousek:2018ago} 
P.~Jiroušek and A.~Vikman,
JCAP {\bf 1904}, 004 (2019),
[arXiv:1811.09547 [gr-qc]].

\bibitem{Gorji:2019ttx} 
M.~A.~Gorji, S.~Mukohyama and H.~Firouzjahi,
JCAP {\bf 1905}, no. 05, 019 (2019),
[arXiv:1903.04845 [gr-qc]].


\bibitem{Chamseddine:2014vna}
A.~H.~Chamseddine, V.~Mukhanov and A.~Vikman,
JCAP {\bf 1406}, 017 (2014),
[arXiv:1403.3961 [astro-ph.CO]].

\bibitem{Mirzagholi:2014ifa} 
L.~Mirzagholi and A.~Vikman,
JCAP {\bf 1506}, 028 (2015), 
[arXiv:1412.7136 [gr-qc]].


\bibitem{Ramazanov:2016xhp} 
S.~Ramazanov, F.~Arroja, M.~Celoria, S.~Matarrese and L.~Pilo,
JHEP {\bf 1606}, 020 (2016), 
[arXiv:1601.05405 [hep-th]].


\bibitem{Ijjas:2016pad} 
A.~Ijjas, J.~Ripley and P.~J.~Steinhardt,
Phys.\ Lett.\ B {\bf 760}, 132 (2016),
[arXiv:1604.08586 [gr-qc]].

\bibitem{Firouzjahi:2017txv} 
H.~Firouzjahi, M.~A.~Gorji and S.~A.~Hosseini Mansoori,
JCAP {\bf 1707}, 031 (2017), 
[arXiv:1703.02923 [hep-th]].



\bibitem{Hirano:2017zox} 
S.~Hirano, S.~Nishi and T.~Kobayashi,
JCAP {\bf 1707}, no. 07, 009 (2017), 
[arXiv:1704.06031 [gr-qc]].

\bibitem{Zheng:2017qfs} 
Y.~Zheng, L.~Shen, Y.~Mou and M.~Li,
JCAP {\bf 1708}, no. 08, 040 (2017), 
[arXiv:1704.06834 [gr-qc]].

\bibitem{Gorji:2017cai} 
M.~A.~Gorji, S.~A.~Hosseini Mansoori and H.~Firouzjahi,
JCAP {\bf 1801}, no. 01, 020 (2018), 
[arXiv:1709.09988 [astro-ph.CO]].

\bibitem{Langlois:2018jdg} 
D.~Langlois, M.~Mancarella, K.~Noui and F.~Vernizzi,
JCAP {\bf 1902}, 036 (2019)
doi:10.1088/1475-7516/2019/02/036
[arXiv:1802.03394 [gr-qc]].

\bibitem{Myrzakulov:2015kda} 
R.~Myrzakulov, L.~Sebastiani, S.~Vagnozzi and S.~Zerbini,
Class.\ Quant.\ Grav.\  {\bf 33}, no. 12, 125005 (2016),
[arXiv:1510.02284 [gr-qc]].

\bibitem{Oikonomou:2016fxb} 
V.~K.~Oikonomou,
Int.\ J.\ Mod.\ Phys.\ D {\bf 25}, no. 07, 1650078 (2016),
[arXiv:1605.00583 [gr-qc]].

\bibitem{Chen:2017ify} 
C.~Y.~Chen, M.~Bouhmadi-López and P.~Chen,
Eur.\ Phys.\ J.\ C {\bf 78}, no. 1, 59 (2018), 
[arXiv:1710.10638 [gr-qc]].

\bibitem{BenAchour:2017ivq} 
J.~Ben Achour, F.~Lamy, H.~Liu and K.~Noui,
JCAP {\bf 1805}, 072 (2018)
doi:10.1088/1475-7516/2018/05/072
[arXiv:1712.03876 [gr-qc]].

\bibitem{Li:2018uwg} 
X.~z.~Li, X.~h.~Zhai and P.~Li,
arXiv:1807.08270 [gr-qc].

\bibitem{Nashed:2018qag} 
G.~G.~L.~Nashed, W.~El Hanafy and K.~Bamba,
JCAP {\bf 1901}, no. 01, 058 (2019), 
[arXiv:1809.02289 [gr-qc]].

\bibitem{Sheykhi:2019gvk} 
A.~Sheykhi and S.~Grunau,
arXiv:1911.13072 [gr-qc].

\bibitem{Herdeiro:2014goa} 
C.~A.~R.~Herdeiro and E.~Radu,
Phys.\ Rev.\ Lett.\  {\bf 112}, 221101 (2014)
doi:10.1103/PhysRevLett.112.221101
[arXiv:1403.2757 [gr-qc]].

\bibitem{Sotiriou:2015pka} 
T.~P.~Sotiriou,
Class.\ Quant.\ Grav.\  {\bf 32}, no. 21, 214002 (2015)
doi:10.1088/0264-9381/32/21/214002
[arXiv:1505.00248 [gr-qc]].





\bibitem{Babichev:2013cya} 
E.~Babichev and C.~Charmousis,
JHEP {\bf 1408}, 106 (2014)
doi:10.1007/JHEP08(2014)106
[arXiv:1312.3204 [gr-qc]].

\bibitem{Sotiriou:2014pfa} 
T.~P.~Sotiriou and S.~Y.~Zhou,
Phys.\ Rev.\ D {\bf 90}, 124063 (2014)
doi:10.1103/PhysRevD.90.124063
[arXiv:1408.1698 [gr-qc]].

\bibitem{Babichev:2016rlq} 
E.~Babichev, C.~Charmousis and A.~Lehébel,
Class.\ Quant.\ Grav.\  {\bf 33}, no. 15, 154002 (2016)
doi:10.1088/0264-9381/33/15/154002
[arXiv:1604.06402 [gr-qc]].

\bibitem{Minamitsuji:2018vuw} 
M.~Minamitsuji and H.~Motohashi,
Phys.\ Rev.\ D {\bf 98}, no. 8, 084027 (2018)
doi:10.1103/PhysRevD.98.084027
[arXiv:1809.06611 [gr-qc]].

\bibitem{BenAchour:2018dap} 
J.~Ben Achour and H.~Liu,
Phys.\ Rev.\ D {\bf 99}, no. 6, 064042 (2019)
doi:10.1103/PhysRevD.99.064042
[arXiv:1811.05369 [gr-qc]].

\bibitem{BenAchour:2019fdf} 
J.~Ben Achour, H.~Liu and S.~Mukohyama,
JCAP {\bf 2002}, no. 02, 023 (2020)
doi:10.1088/1475-7516/2020/02/023
[arXiv:1910.11017 [gr-qc]].

\bibitem{Motohashi:2019ymr} 
H.~Motohashi and S.~Mukohyama,
JCAP {\bf 2001}, no. 01, 030 (2020)
doi:10.1088/1475-7516/2020/01/030
[arXiv:1912.00378 [gr-qc]].

\bibitem{mukohyama-nogo} 
K.~Izumi and S.~Mukohyama,
Phys.\ Rev.\ D {\bf 81}, 044008 (2010)
doi:10.1103/PhysRevD.81.044008
[arXiv:0911.1814 [hep-th]].


\bibitem{Blas:2009yd} 
D.~Blas, O.~Pujolas and S.~Sibiryakov,
JHEP {\bf 0910}, 029 (2009)
doi:10.1088/1126-6708/2009/10/029
[arXiv:0906.3046 [hep-th]].

\bibitem{Barausse:2011pu} 
E.~Barausse, T.~Jacobson and T.~P.~Sotiriou,
Phys.\ Rev.\ D {\bf 83}, 124043 (2011)
doi:10.1103/PhysRevD.83.124043
[arXiv:1104.2889 [gr-qc]].

\bibitem{Baumgarte:2010ndz} 
T.~W.~Baumgarte and S.~L.~Shapiro,
doi:10.1017/CBO9781139193344

\bibitem{Zumalacarregui:2013pma} 
M.~Zumalacárregui and J.~García-Bellido,
Phys.\ Rev.\ D {\bf 89}, 064046 (2014)
doi:10.1103/PhysRevD.89.064046
[arXiv:1308.4685 [gr-qc]].


\end{thebibliography}
\end{document}